\documentclass[letterpaper,12pt]{article}
\usepackage[utf8]{inputenc}
\usepackage[inner=2.5cm,outer=2.5cm]{geometry}
\usepackage[english,activeacute]{babel}
\usepackage{amssymb,amsmath,alltt}
\usepackage{amsfonts}
\usepackage{enumitem}
\usepackage{authblk}
\usepackage{soul}
\usepackage{hyperref}

\usepackage[title]{appendix}
\usepackage{xcolor}
\usepackage[section]{placeins}

\definecolor{red2}{RGB}{250,0,0}
\definecolor{blue2}{RGB}{0,0,180}
\definecolor{green2}{RGB}{0,180,0}
\definecolor{purple2}{RGB}{120,0,120}
\definecolor{grey2}{gray}{0.6}

\newcommand{\R}{\mathbb{R}}

\newcommand{\la}{\langle}
\newcommand{\ra}{\rangle}
\newcommand{\curl}{\textrm{curl}}
\newcommand{\eps}{\varepsilon}
\renewcommand{\div}{\textrm{div}}

\newtheorem{thm}{Theorem}[section]
\newtheorem{rem}[thm]{Remark}

\usepackage{background}
\backgroundsetup{contents={}}

\title{Regions without flux surfaces of given class for\\magnetic fields in toroidal geometry}
\author[1,2]{N.Kallinikos}
\author[3]{R.S.MacKay}
\author[4]{D.Martínez-del-Río\footnote{david.martinez-del-rio@warwick.ac.uk}}
\affil[1]{Department of Mathematics, University of Western Macedonia, 52100 Kastoria, Greece}
\affil[2]{Department of Physics, International Hellenic University, 65404 Kavala, Greece}
\affil[3]{Mathematics Institute, University of Warwick, Coventry CV4 7AL, UK}
\affil[4]{Department of Statistics, University of Warwick, Coventry CV4 7AL, UK}

\date{}

\begin{document}

\maketitle

\begin{abstract}
A Converse KAM method for 3D vector fields, establishing regions through which pass no invariant 2-tori transverse to a given direction field, is tested on some helical perturbations of an axisymmetric magnetic field in toroidal geometry. It finds regions corresponding to magnetic islands and chaos for the fieldline flow. Minimization of these regions is proposed as a tool to help in the design of plasma confinement devices of tokamak and stellarator type.
\end{abstract}


\section{Introduction}
KAM theory provides sufficient conditions for the existence of invariant tori in Hamiltonian systems.  In particular, many invariant tori persist from generic integrable Hamiltonian systems under smooth and small enough perturbations (for a semi-popular introduction, see \cite{dumas2014kam}). 
Nevertheless, it is still hard work to prove existence of a realistic fraction of the tori that are suggested to exist by numerical simulation, e.g.~\cite{figueras2017rigorous}. 

An alternative approach is to determine regions through which  no invariant tori of given class pass. Termed \textit{Converse KAM theory} \cite{mackay1985converse,mackay1989criterion}, it is much easier to implement than KAM theory and gives close to  optimal conclusions without excessive computation.

The present work is an application to magnetic fields of Converse KAM theory, as extended in \cite{mackay2018finding} to allow more general classes of tori than the earlier references and to treat 3D vector fields rather than positive-definite Lagrangian systems. It follows the main points of the implementation presented in \cite{kallinikos2022regions}, which was for the planar circular restricted three-body problem on level sets of the Jacobi constant. 

The case of study in this work is magnetic fields in toroidal configurations, in particular the identification of regions through which pass no invariant tori (flux surfaces) of a given class.  We define a class of tori by specifying a direction field almost everywhere and asking for tori that are transverse to that direction field. The principal choice of direction field is the gradient of a suitable notion of distance from a closed fieldline (magnetic axis), with respect to a chosen metric.


Our method uses the magnetic flux-form as principal representation of a magnetic field.  Given a vector field $B$ preserving a volume-form $\Omega$, its flux-form is $\beta = i_B\Omega$. The integral $\int_S \beta$ over any surface $S$ (with boundary allowed) represents the magnetic flux through that surface. Appendix \ref{app:pedagogy} presents a summary of relevant background. For a tutorial about the use of differential forms in plasma physics, see \cite{mackay2020differential}.

In Section \ref{sec:magfields}, we introduce the magnetic fields to be studied in this paper. In Section \ref{sec:cKAM}, we explain how to apply the Converse KAM method to magnetic fields. In Section \ref{sec:results}, we present the results of numerical implementation of the method on the chosen fields. Section \ref{sec:conc} discusses the results. Finally, three appendices give pedagogical introductions to some of the mathematics.

\section{Toroidally helical magnetic fields}
\label{sec:magfields}
The magnetic fields that we choose to illustrate the Converse KAM method here are perturbations of a circular tokamak field by helical modes, based on \cite{kallinikos2014integrable}.  They have the advantages that:
\begin{enumerate}
\item there is an explicit magnetic axis and an easily specified class of tori that surround it;
\item with a single helical mode, the field is still integrable, but has a computable island; the invariant tori outside the island all belong to the chosen class and none of those inside the island do, so the method can be tested on its ability to detect the island;
\item with more than one helical mode, the field can be expected to have the typical mix of invariant tori of the original class, islands, and chaos, so the method can be tested on such cases;
\item they show how to handle fields presented in non-trivial coordinate systems, which is the typical case for tokamak and stellarator fields.
\end{enumerate}

Given a coordinate system $(x^1,x^2,x^3)$,  we will express a magnetic field $B$ in terms of its contravariant components $B^{i}$, rather than its physical ones; they differ by length factors (see Appendix~\ref{app:curv} for a summary about components of vector fields in curvilinear coordinates). An advantage is that the equations of motion for fieldline flow are just $\dot{x}^i = B^{i}(x)$; here ``time'' is to be understood along the magnetic field lines.

Our fields are simplest described and treated in an adapted toroidal coordinate system $(\psi,\vartheta,\phi)$,  
which is a variant of the standard toroidal coordinates $(r,\theta,\phi)$. It is not straightforward to describe the coordinate system, and it needs first considerations on $B$, in particular its toroidal component $B^\phi$, but we do it in the next subsection. It might seem demanding, but working in a non-trivial coordinate system is likely to be  part of any application to realistic fields.

\begin{figure}[ht!]
 \centering  
   \includegraphics[width=0.5\linewidth]{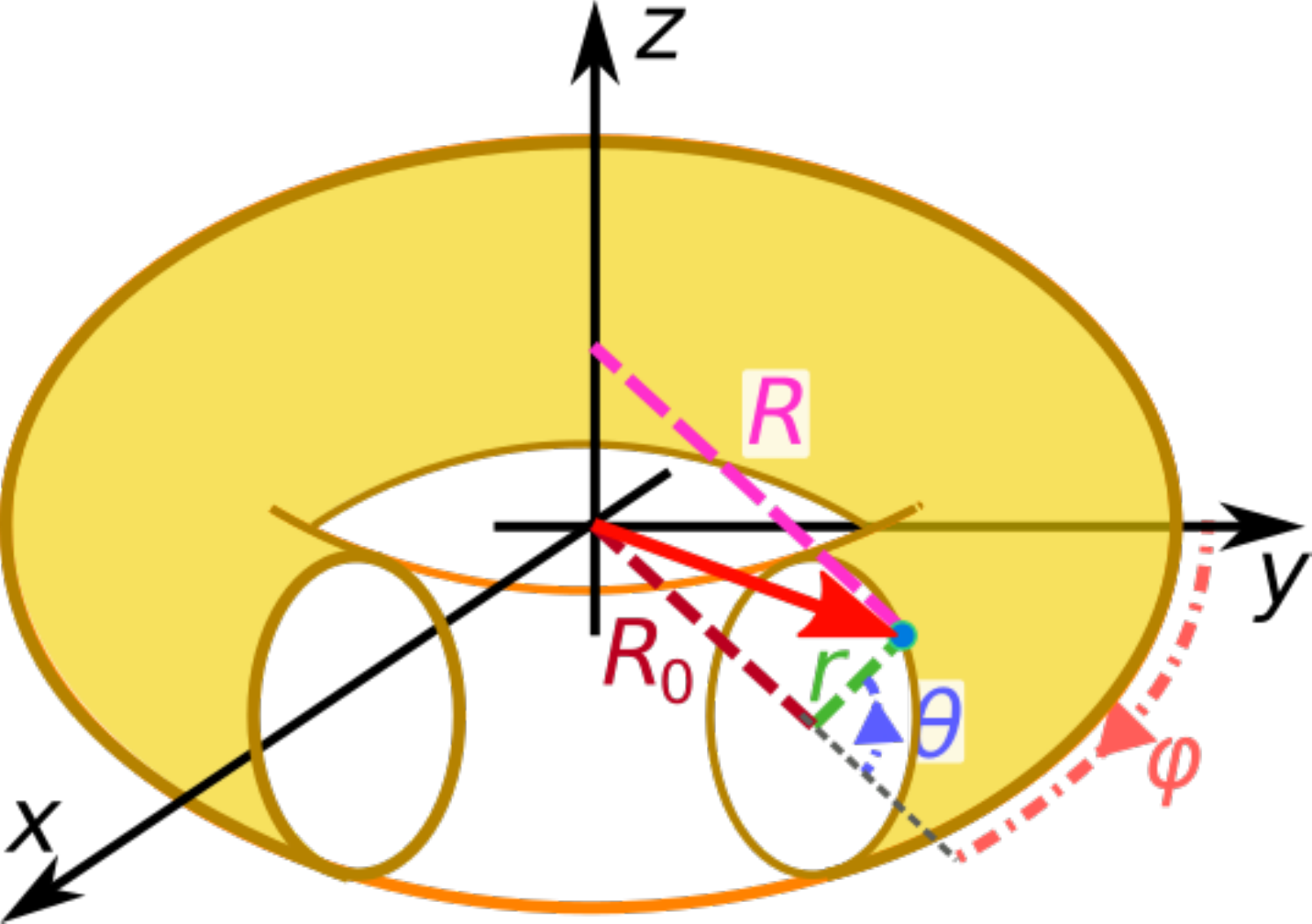}
   \caption{Toroidal coordinates.} 
   \label{Fig_toroidal}
\end{figure}

First we recall the standard toroidal coordinates
$(r,\theta,\phi)$. They are related to Cartesian coordinates $(x,y,z)$ through 
$$x = R \sin\phi,\ y = R\cos \phi,\ z = r\sin \theta,$$
where
$$R= R_0 + r\cos\theta,$$
for some $R_0>0$ and $0\le r<R_0$. As shown in Figure \ref{Fig_toroidal}, $R_0$ is the radius of the magnetic axis and $R$ represents the cylindrical radius relative to the $z$-axis. In these coordinates, the metric tensor is represented by the matrix $\text{diag}(1,r^2,R^2)$.

\subsection{Construction of adapted coordinates}
Following \cite{kallinikos2014integrable}, we introduce coordinates $(\psi,\vartheta)$ to make the restriction $\beta_T$ of the magnetic flux-form $\beta$ to a poloidal section ($\phi=$ constant) take the form
\begin{equation}
\label{eq:tflux2}
\beta_{\text T}=d\psi \wedge d\vartheta,
\end{equation}
where $\wedge$ denotes the exterior product of differential forms (see \cite{mackay2020differential} for a tutorial).
First, we define $\psi$ as the toroidal magnetic flux across the poloidal disk of radius $r$ about a point of the magnetic axis, divided by $2\pi$. Thus, integrating
\begin{equation}
\label{eq:tflux1}
\beta_{\text T}=rR B^\phi\, dr \wedge d\theta,
\end{equation}
over a poloidal disk of radius $r$ yields $\psi$. Then $\vartheta$ can be constructed by equating (\ref{eq:tflux2}) and (\ref{eq:tflux1}), i.e., the condition
\begin{equation}
\label{eq:concor}
rR B^\phi\, dr \wedge d\theta=d\psi \wedge d\vartheta.
\end{equation}

Therefore the transformation $(r,\theta)\longmapsto(\psi,\vartheta)$ basically relies on the toroidal component $B^\phi$. We choose our magnetic fields to all have
\begin{equation}
\label{eq:tB}
B^\phi = \frac{B_0R_0}{R^2}
\end{equation}
with $B_0>0$. This is a simple but realistic form for $B^\phi$, corresponding to external poloidal current $2\pi R_0 B_0/\mu_0$. Hence we arrive at \cite{kallinikos2014integrable, abdullaev2014}
\begin{align}
\begin{split}
&\psi= B_0R_0\left(R_0-\sqrt{R_0^2-r^2}\right)\\
&\tan \frac{\vartheta}{2} = \sqrt{\frac{R_0-r}{R_0+r}}\tan\frac{\theta}{2}.
\end{split}
\end{align}
Note that $\psi \sim B_0 r^2/2$ as $r/R_0 \longrightarrow 0$, $\psi$ is restricted to non-negative values less than $B_0R_0^2$, and has a coordinate singularity at $0$.

The magnetic flux-form $\beta$ plays a key role in the application of the Converse KAM method, so it is useful to simplify its expression by suitable coordinates. Here, in particular, we focused on the restriction $\beta_{\text T}$ of $\beta$ to poloidal sections, because we will see that for our examples we can implement the method using only $\beta_{\text T}$. For more general implementation though, it is essential to use the full flux-form $\beta$.

\begin{rem}\normalfont
Another way of thinking of $(\psi,\vartheta)$ and $\beta_{\text T}$ is related to the standard Hamiltonian formulation of magnetic fields. 
The latter typically uses $\phi$ as time along the field lines, assuming $B^\phi\neq0$. Field line flow can be written then as a time-dependent Hamiltonian system with Hamiltonian function $H=-\,A_\phi$ ($A$ being the vector potential, see below) and symplectic form $\omega=\beta_{\text T}$. Thus, bringing (\ref{eq:tflux1}) to the form (\ref{eq:tflux2}) simply amounts to finding canonical coordinates $(\psi,\vartheta)$ for $\omega$. The Hamiltonian treatment though is neither necessary nor simpler for either Converse KAM or the helical fields we use. The flux-form $\beta$ is key instead.
\end{rem}

\subsection{Magnetic fields studied}
To enforce volume-preservation by the fieldline flow, we specify $B$ as the curl of a vector potential $A$.  In terms of the covariant components of $A$, the contravariant components of $B$ are given by
$$B^{i} = \frac{1}{\sqrt{|g|}} \epsilon^{ijk} \partial_j A_k,$$
where $\epsilon$ is the Levi-Civita symbol and $|g|$ is the determinant of the matrix $g$ representing the metric tensor, $ds^2 = g_{ij} dx^i dx^j$. In our adapted toroidal coordinates, the volume factor $\sqrt{|g|}$ is $ 1/B^\phi=R^2/(B_0R_0).$  This can be shown without finding $g$, by computing the volume-form $\Omega = dx\wedge dy\wedge dz=rR \ dr \wedge d\theta \wedge d\phi$ and using (\ref{eq:concor})-(\ref{eq:tB}) for the toroidal flux.

We take a vector potential with helical modes introduced in its toroidal component, of the form (in covariant components)
\begin{align}
\label{eq:potential}
\begin{split}
A_\psi &= 0\\
A_\vartheta &= \psi \\
A_\phi &= -[w_1\psi+w_2\psi^2 + \sum_{m,n} \eps_{mn} \psi^{m/2} f_{mn}(\psi) \cos(m\vartheta-n\phi + \zeta_{mn})],
\end{split}
\end{align}
where $w_1\in\R$, $w_2\ne 0$, $m,n$ are integers with $m\ge 2$, $f_{mn}$ are smooth functions and $\zeta_{mn}$ arbitrary phases.  The factor $\psi^{m/2}$ is to make the resulting vector potential smooth at $\psi=0$ (this was not done in \cite{kallinikos2014integrable} but its authors were interested there only in the neighbourhoods of islands).
The coefficient $w_2$ produces shear.  Extension to examples with change of sign of shear could be achieved by adding a term $w_3 \psi^3$; that would be a good next test case for the method, because in contrast to \cite{mackay1989criterion} the method here does not require shear, but we leave it for future work.

The vector potential (\ref{eq:potential}) gives rise to the magnetic field $B=(B_0R_0/R^2) V$ where the components of the auxiliary vector field $V$ are
\begin{align}
\label{eq:fields}
\begin{split}
V^\psi &= \sum_{m,n} m\eps_{mn}\psi^{m/2} f_{mn}(\psi)\sin(m\vartheta-n\phi + \zeta_{mn})\\
V^\vartheta &= w_1 + 2 w_2 \psi + \sum_{m,n} \eps_{mn} \psi^{m/2-1} \left[\tfrac{m}{2}f_{mn}(\psi)+\psi f_{mn}'(\psi)\right]\cos(m\vartheta-n\phi + \zeta_{mn})\\
V^\phi &= 1 .
\end{split}
\end{align}
The cylindrical radius $R$ occurring in the conversion from $V$ to $B$ can be expressed in our adapted coordinates via
$$R = \frac{R_0^2-r^2}{R_0-r\cos\vartheta}$$
with $$r = \sqrt{2\frac{\psi}{B_0}-\frac{\psi^2}{B_0^2R_0^2}},$$
but we can avoid the conversion by applying the Converse KAM method to $V$ rather than $B$, as will be explained.

Because we take $m\ge 2$, the fields all have $\psi=0$ as a closed fieldline, as claimed, which we call the {\em magnetic axis}.

We define the {\em principal class} of tori to be the differentiable tori that are transverse to $\nabla \psi$.
For example, with no helical modes the field is integrable with integral $\psi$  and the invariant tori $\psi=$ constant belong to the principal class.  So do all $C^1$-small deformations of them.
Specifying $\nabla \psi$ entails a choice of Riemannian metric, but there is no need to use the Euclidean one, especially as in the adapted toroidal coordinates its computation would add extra work.  It is preferable to choose a metric so that $\nabla \psi$ is in the same direction as $\partial_\psi$ (see Appendix \ref{app:curv} for the distinction; in particular, this fails for the Euclidean metric:~the adapted toroidal coordinates are not orthogonal).  Then the principal class of tori consists of the graphs of $\psi$ as a differentiable function of $(\vartheta,\phi)$.  We keep the more general specification $\nabla \psi$, however, for flexibility.

The last ingredient to describe is the full magnetic flux-form $\beta$ (as opposed to just its restriction $\beta_{\text T}$ to poloidal sections).  This is defined by $\beta = i_B \Omega$ where $\Omega$ is the volume form, or equivalently by $\beta = dA^\flat$, where $$A^\flat = A_\psi d\psi + A_\vartheta d\vartheta + A_\phi d\phi$$ (indeed, it is better to think of the vector potential $A$ as a 1-form potential $A^\flat$ for $\beta$).  Thus 
$$\beta = V^\psi d\vartheta \wedge d\phi + V^\vartheta d\phi \wedge d\psi + V^\phi d\psi \wedge d\vartheta.$$
Because $V^\phi=1$, we see that restricted to a poloidal section, $\beta = \beta_{\text T} = d\psi \wedge d\vartheta$, as claimed earlier.

\subsubsection{Single helical mode}
\label{subsec:integrable}
A nice feature of our chosen form of field is that with a single helical mode, the field is still integrable \cite{kallinikos2014integrable}.  Indeed, it has the invariant (i.e., integral of motion) 
\begin{equation}
\Psi = -n\psi-mA_\phi\,.
\label{Psi}
\end{equation}
This can be checked directly. Alternatively, it can be derived from the symmetry $u = n\partial_\vartheta+m\partial_\phi$, as follows.  The vector field $u$ preserves the components $A_\psi,A_\vartheta,A_\phi$ in (\ref{eq:potential}), and therefore $A^\flat$, i.e., $L_uA^\flat=0$. Thus, the rate of change of $u\cdot A$ along $B$ is
$$L_B(u\cdot A)=i_B d i_u A^\flat = i_B (L_u-i_ud) A^\flat = i_B(L_uA^\flat - i_ui_B\Omega) = 0,$$
meaning $\Psi=-u\cdot A$ is conserved by $B$. 
This result holds not only for fields with a single helical mode but also for any field with potential (\ref{eq:potential}) in which $A_\phi$ is a function of only $\psi$ and a single combination $m\vartheta-n\phi$ of the angle variables.
Note that although $\beta$ is invariant under $u$, since $L_u\beta=dL_uA^\flat=0$, the magnetic field $B$ itself is not, as $u$ is not volume-preserving.

\begin{figure}[ht!]
 \centering  
    \includegraphics[height=7.3cm, trim={0 0 0 0},clip]{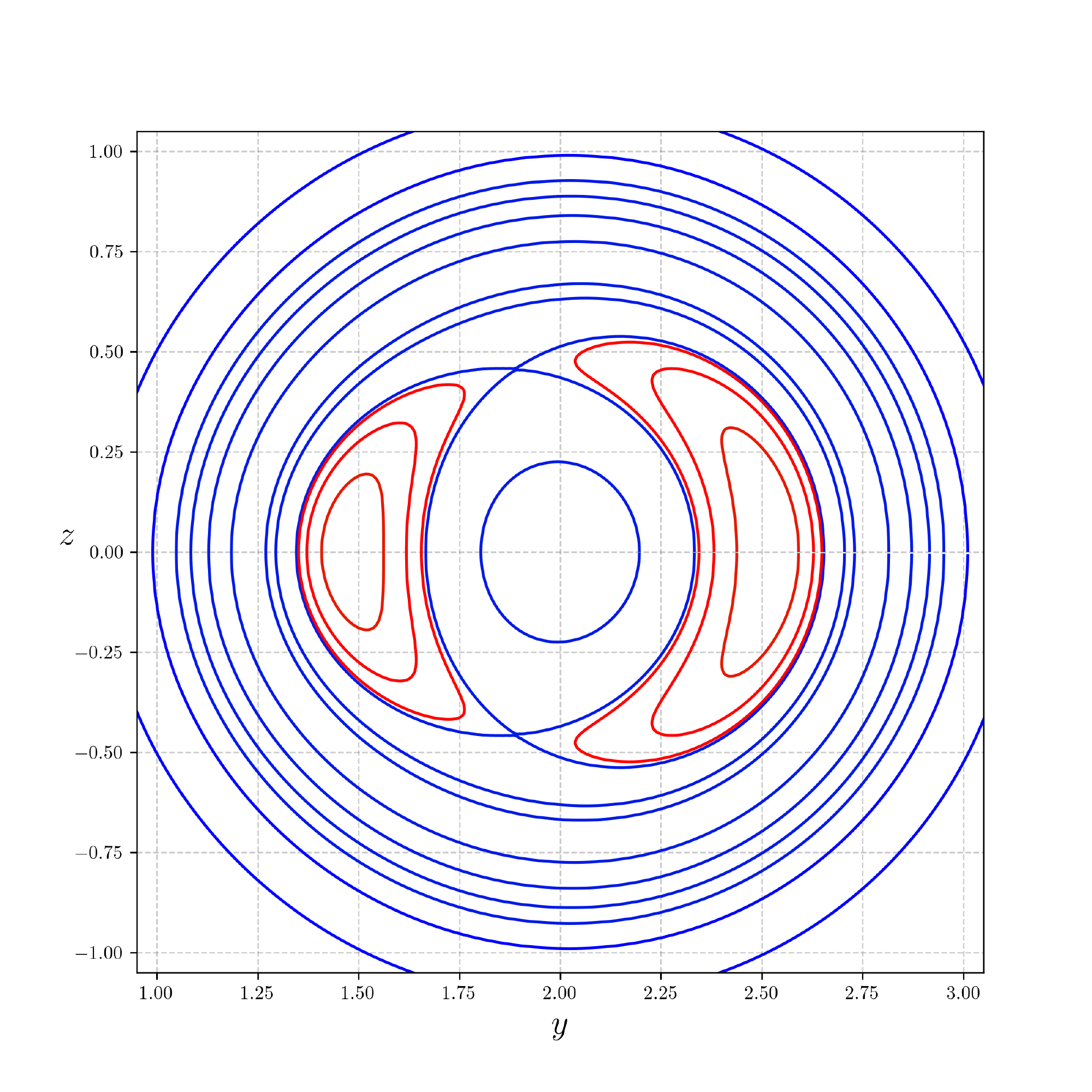} 
    \includegraphics[height=7.3cm, trim={0 0 0 0},clip]{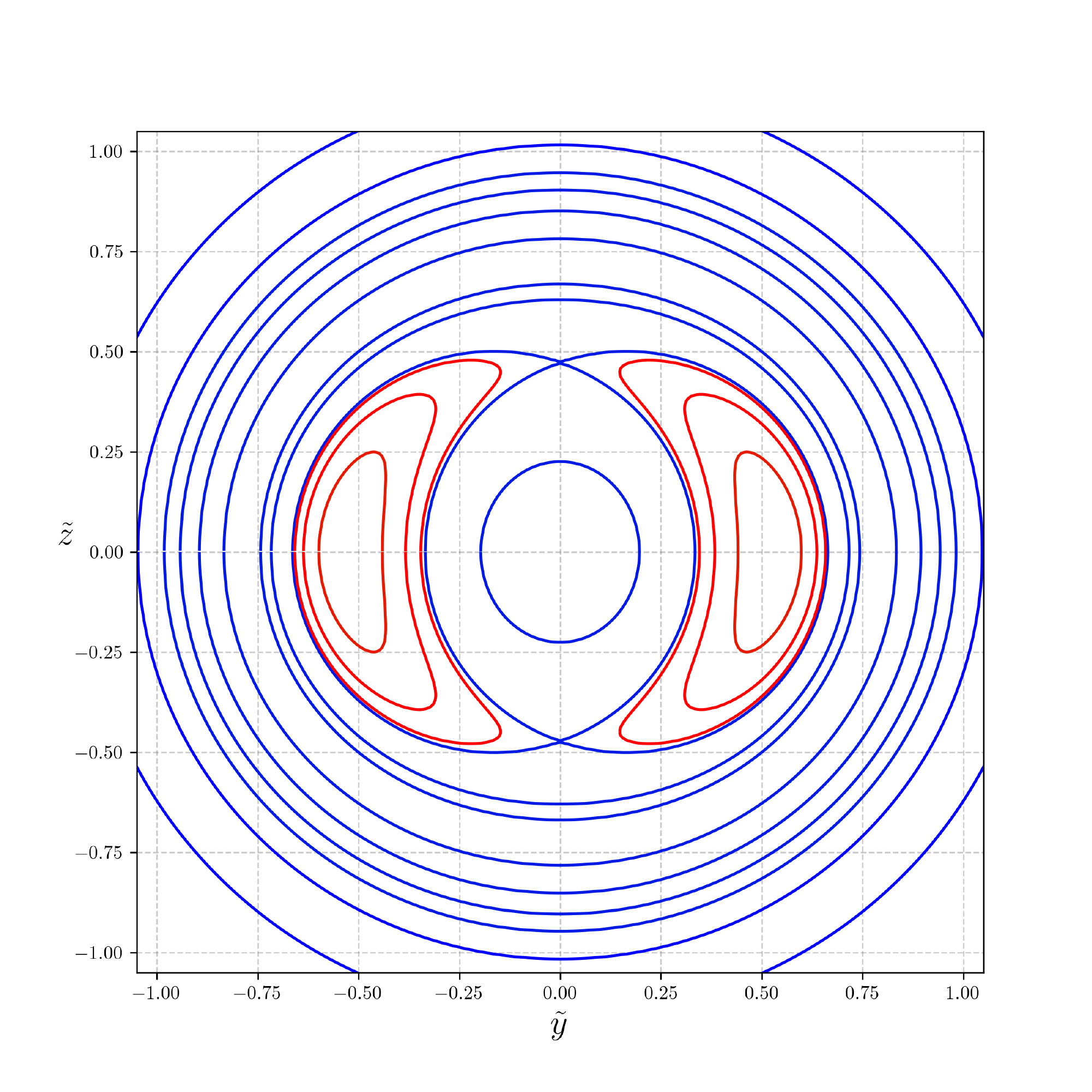}
   \caption{Level sets of $\Psi$ on the poloidal section $\phi=0$ for $(m,n)=(2,1)$ and standard values given by (\ref{eq:standard_values}), in Cartesian (left) and symplectic (right) coordinates.}
   \label{Fig_coords_comp_2_1}
\end{figure}

In general, the integral gives rise to a family of invariant tori of the principal class and (if the signs are appropriate) a family that foliate an island. See Figure \ref{Fig_coords_comp_2_1} for an example on a poloidal section. 

To help orient the reader, we plotted this figure first in Cartesian coordinates, but in future we will just plot in ``symplectic'' coordinates
\begin{align}
\label{eq:sym_cords}
\begin{split}
\tilde{y}& = \sqrt{2\psi/B_0}\cos\vartheta\\
\tilde{z}& = \sqrt{2\psi/B_0}\sin{\vartheta}
\end{split}
\end{align}
on the poloidal section $\phi=0$. Near the magnetic axis this is a small distortion (especially for large aspect ratio $r/R_0 \longrightarrow 0$) of the true $yz$-plane $x=0$, but with area equal to toroidal flux and the magnetic axis shifted to the origin.

In particular, for $m=2$ and $f(\psi) = f_0 + f_1\psi$, completing the square shows that the tori are the components of the sets where
\begin{equation}
\left(\psi-\frac{n/2-w_1-\eps f_0\cos\zeta}{2(w_2+\eps f_1 \cos\zeta)}\right)^2 = \frac{\Psi}{2(w_2+\eps f_1\cos\zeta)} + \frac{(n/2-w_1-\eps f_0\cos\zeta)^2}{4(w_2+\eps f_1\cos\zeta)^2},
\label{eq:tori}
\end{equation}
with $\zeta = 2\vartheta-n\phi+\zeta_{2n}$. These are graphs of $\psi$ as a function of $(\vartheta,\phi)$ if and only if the righthand side is everywhere positive. Supposing $w_2>0$, $w_1<n/2$,  $(n/2-w_1)f_1+2w_2f_0 \ne 0$ and $\eps>0$ is small enough, we obtain the island explicitly as the set where
\begin{equation}\Psi \le -\frac12 \frac{(n/2-w_1-\eps f_0)^2}{w_2+\eps f_1}.
\label{eq:island}
\end{equation}
The formula for other combinations of signs can be obtained if desired, but if $|(n/2-w_1)f_1+2w_2f_0| <\eps$ there is a more complicated island with four critical points.
Note that the tori outside the island might not all be transverse to $\nabla \psi$ if the metric is not well chosen and $\eps$ is not small, but they are transverse to $\partial_\psi$.  This explains our preference for a metric such that $\nabla \psi$ is in the direction of $\partial_\psi$.

\section{Converse KAM for magnetic fields}
\label{sec:cKAM}

The basic idea of Converse KAM methods is to consider how infinitesimal displacements (``tangent vectors'' in mathematical terminology) rotate under a flow. If a 3D flow has an invariant orientable surface of a given class then it prevents infinitesimal displacements from rotating from one side of it to the other. So if an infinitesimal displacement from some trajectory rotates incompatibly with this restriction then there is no invariant surface containing that trajectory.

To make this precise, we have to specify a class of surfaces, in particular tori, and make clear what qualifies as rotating from one side to the other for all candidate tori in this class.  We achieve these by choosing a ``direction field'' $\xi$ and a 1-form $\lambda$, both to be explained below, and using the magnetic flux-form $\beta$.  In the special case of fields with ``stellarator symmetry'' we explain how to streamline the method for symmetric trajectories.

Converse KAM theory in continuous time was first developed for Hamiltonian systems. It is standard knowledge that magnetic fields can be regarded as Hamiltonian systems (Appendix~\ref{app:Hamsys} describes the way we prefer to do this), but it is more straightforward to work directly with the magnetic flux-form.

\subsection{Direction field}

For a vector field $B$ on an oriented 3D space, the Converse KAM method of \cite{mackay2018finding} eliminates regions through which pass no invariant tori of $B$ transverse to a given 1D foliation. A continuous choice of orientation can be assigned to the leaves of the foliation and thus a continuous choice of non-zero vectors $\xi$ tangent to the foliation can be made, indicating the orientation. Because only the direction matters, not the magnitude, we call $\xi$ a {\em direction field} (in standard differential-geometric terminology, $\xi$ is the distribution associated to the foliation).

As presented in the previous section, we  choose direction field for the principal class of tori in our examples to be $\nabla\psi$ with respect to some metric (which need not be the Euclidean one). In practise, we chose the metric to make $\xi$ be in the direction of $\partial_\psi$, so that we can be sure of the classification of tori for integrable fields with one helical mode.  We present the method for general $\nabla \psi$, however, for compatibility with \cite{kallinikos2022regions} and potential applications to include island tori where we'd replace $\psi$ by $\Psi$ of equation (\ref{Psi}), and to guiding-centre motion.

An important extension is required, however, to cater for classes of tori around a magnetic axis.  Namely, we allow the direction field $\xi$ to have zeroes.  This is the case for $\nabla \psi$ on the magnetic axis, for example.  Note that if a torus is transverse to $\xi$ then a fortiori it does not intersect the zero-set of $\xi$.


\subsection{Nonexistence condition}

The method of \cite{mackay2018finding} gives a sufficient condition for the non-existence of invariant tori of a 3D vector field through a given point, transverse to a direction field $\xi$. We describe here its adaptation to magnetic fields $B$.
We will assume $B$ is nowhere zero in the domain of interest; equivalently, the kernel of $\beta$ is one-dimensional at every point.

Before we start, for any positive function $f$, the vector field $V=B/f$ has the same invariant tori as $B$.  So it is a good idea to choose a function $f$ to simplify the expression of $V$.  See the previous section, for example.  In general, $V$ no longer preserves the same volume-form $\Omega$ as $B$ but it preserves the related volume-form $f\Omega$.  Also the important relation $i_B\beta = 0$ is inherited by $V$:~$i_V\beta=0$.
We will treat $B$ in what follows, but one should bear in mind this possibly useful pre-processing.

Given an initial point $s_0=s(0)$ in 3D space, take initial tangent vector $\eta_{s_0} = \xi_{s_0}$. For $t$ positive or negative, compute the propagation of $s(t)$ and $\eta_{s(t)}$ under the dynamics $\dot{s} = B(s)$ and the linearised dynamics $\dot{\eta}_s = DB_s\,\eta_s$, respectively. If there is an invariant torus $\mathcal{T}$ passing though $s_0$ that is transverse to $\xi$, then $\eta_{s(t)}$ must stay on the same side of $\mathcal{T}$ for all $t$. In particular, $\eta_s$ is never of the form $c_1 \xi_s + c_2B_s$ with $c_1<0$.  Checking this condition can be broken down into two steps:
\begin{enumerate}[label=(\roman*)]
\item examine if $(\eta_s,\xi_s,B_s)$ pass through a case of linear dependence; 
\item if so, examine the sign of $c_1$.  
\end{enumerate}
If one finds a time at which the stated vectors are linearly dependent with $c_1<0$ then the given trajectory does not lie on any invariant torus transverse to the given field $\xi$.

The conditions (i) and (ii) are stopping criteria for the integration of $s(t)$ and $\eta_{s(t)}$. To detect them, we follow the ``general'' formulation of \cite{kallinikos2022regions}. In the present context, this uses the magnetic flux-form $\beta$ for (i) and a 1-form $\lambda$ for (ii) such that $\lambda({B}) = 0$ and $\lambda(\xi) > 0$ (except on zeroes of $\xi$). The reason that $\beta$ suffices here (instead of the symplectic form on energy levels used in \cite{kallinikos2022regions}) is that $B$ belongs to the kernel of $\beta$, hence the triple product of $(\eta_s,\xi_s,B_s)$ using the standard volume-form $\Omega=|B|^{-2}B^\flat\wedge\beta$ reduces to $\beta(\eta_s,\xi_s)$.  Therefore, the general formulation translated to magnetic fields says that
\begin{thm}\label{thm:gen}
Given initial conditions $s_0=s(0)$, choose initial vector $\eta_{s_0}=\xi_{s_0}$. If there is a time $t$ such that at $s=s(t)$: (i) $\beta(\eta_s,\xi_s)$ changes sign, and (ii) $\lambda(\eta_s) < 0$, then there is no invariant torus through $s_0$ transverse to $\xi$. 
\end{thm}
Thus, we can mark $s_0$ (and indeed its whole forward and backward orbit) as a point of the region of nonexistence of the desired class of invariant tori. In other words, we can eliminate $s_0$ (and its orbit) from being on invariant tori of the given class.

To apply this theorem, one needs to choose a 1-form $\lambda$ with the required properties.  There is some freedom here.  
In the case of $\xi = \nabla \psi$ for some choice of metric,
following the steps from  \cite{kallinikos2022regions}, we choose $\lambda = d\psi - k B^\flat$, where $k = B\cdot\nabla \psi/|B|^2 = B^\psi/|B|^2$ and both $B^\flat$ and $|B|^2$ are defined using the chosen metric ($B^\flat$ is the 1-form such that for all vectors $u$, $B^\flat u = B \cdot u$). By construction, $\lambda(B) = 0$ and $\lambda(\xi) > 0$ everywhere except where $\xi$ is parallel to $B$ (by the Cauchy-Schwarz inequality). In our case, $B^\phi >0$ implies that the only places where $\xi$ is parallel to $B$ are where $\xi=0$, i.e., on the magnetic axis.  

For our examples, it is simplest if the chosen metric is diagonal in adapted toroidal coordinates. It is best also if the metric makes dimensional sense. Thus we use
$$ds^2 = \frac{1}{2B_0\psi} d\psi^2 + \frac{2\psi}{B_0}d\vartheta^2 + R_0^2 d\phi^2,$$
which approximates Euclidean metric near the magnetic axis.

In vector calculus notation, the two quantities in Theorem \ref{thm:gen} are expressed as $\beta(\eta,\xi)=\xi\cdot\beta\eta$ and $\lambda(\eta)=\eta\cdot(\nabla \psi-kB)$ for the aforementioned choice for $\lambda$, considering $\beta$ as a matrix (as described in Appendix \ref{app:curv}), and $\cdot$ denoting the dot product with respect to the metric used.  Also, note that $\eta \cdot \nabla \psi = \eta^\psi$, independently of the metric.

Note that if we use $V$ instead of $B$ for our examples then $V^\phi=1$ implies that an initially poloidal vector remains poloidal.  Assuming $\xi$ is chosen to be poloidal, this means firstly, the integration of tangent orbits does not require the $\phi$-component to be represented, and secondly, $\beta$ needs evaluating only on pairs of poloidal vectors, where it has the simple form $d\psi \wedge d\vartheta$. We applied the method to both $B$ and $V$, the first to demonstrate its general applicability, the second to speed up computations.

\subsection{Using stellarator symmetry}
\label{subsec:Stellsym}

The method of the previous subsection has a refinement for systems that admit a time-reversal symmetry, as noted and used in \cite{mackay1989criterion}. In the present context of magnetic field lines, the equivalent is the ``stellarator symmetry'' $\mathcal{R}: (r, \theta , \phi) \longmapsto (r, -\theta , -\phi)$ \cite{dewar1998stellarator}, that translates to $(\psi,\vartheta,\phi) \mapsto (\psi, -\vartheta,-\phi)$. The flow of a magnetic field $B$ has stellarator symmetry if $B$ is $\mathcal{R}$-antisymmetric, i.e., $B\mathcal{R} = -\,d\mathcal{R}\,B$, or equivalently $\tilde{B}_{\tilde{s}}=-B_s$, where $\tilde{B}=d\mathcal{R}\,B$ and $\tilde{s}=\mathcal{R}(s)$. Although this hypothesis narrows the magnetic fields that can be considered, it is a commonly assumed property  in the fusion plasma literature and the design of stellarators.  If the phases $\zeta_{mn}$ are chosen zero then our examples have stellarator symmetry. 

For magnetic fields with stellarator symmetry and initial conditions on a symmetry line (the half-lines of fixed points of $\mathcal{R}$, i.e.~$\theta = 0$ or $\pi$, $\phi=0$ or $\pi$, $r>0$) it is possible to simplify the test and speed up the computation by a factor of at least two, as the backward trajectory is a reflection by $\mathcal{R}$ of the forward one. Only now, we need to choose $\xi$ to be $\mathcal{R}$-symmetric, and to take as initial condition an $\mathcal{R}$-antisymmetric vector $\eta_{s_0}$ (not $\xi_{s_0}$ as in the general formulation) on the symmetry semi-line, independent of $B_{s_0}$.

The two steps of the non-existence condition are then reduced to one, namely $\beta(\eta_s,\xi_s) = 0$ for some $t>0$. This is because if $\eta_{s_0}$ evolves to $\eta_{s(t)}$ for some $t>0$ with $\beta(\eta_s,\xi_s)=0$, then $\eta_s = c_1 \xi_s+c_2B_s$ for some $c_1,c_2$, and $c_1 \ne 0$ since $\eta$ was independent of $B$ and independence is preserved by the evolution. But, by reflection, $\eta_{s_0}$ also evolves backwards in time to $\tilde{\eta}_{\tilde{s}}= -c_1\tilde{\xi}_{\tilde{s}} + c_2\tilde{B}_{\tilde{s}}$. As previously explained, the change in sign of the component along $\xi$ is incompatible with existence of an invariant torus through $s_0$ transverse to $\xi$. 
\begin{thm}\label{thm:sym}
Let $B$ and $\xi$ be $\mathcal{R}$-antisymmetric and $\mathcal{R}$-symmetric, respectively. Given initial conditions $s_0=s(0)$ on the symmetry lines, choose an initial $\mathcal{R}$-antisymmetric vector $\eta_{s_0}$. If there is a time $t$ such that $\beta(\eta_s,\xi_s)$ changes sign at a point $s=s(t)$, then there is no invariant torus through $s_0=s(0)$ transverse to $\xi$. 
\end{thm}

To fix ideas, we choose $\xi = \partial_\psi$ (equivalently, $\psi \partial_\psi$ to make it have a limit on the magnetic axis, as its magnitude doesn't matter) and $\eta_{s_0} = \partial_\vartheta$.  Then they both lie in a poloidal plane and so under the dynamics of $V$, $\eta_s$ remains in a poloidal plane and hence $\beta$ can be simplified to $d\psi \wedge d\vartheta$ again.

\section{Results}
\label{sec:results}

In this section, we apply the Converse KAM method laid out in Section \ref{sec:cKAM} to integrable and non-integrable cases of magnetic fields of the type described in Section \ref{sec:magfields}. In particular, we apply the method to find regions without invariant tori (that is, flux surfaces) transverse to the $\psi$-direction. Eliminated from being on such tori, they reveal magnetic islands and chaotic regions.

More specifically, Theorem \ref{thm:gen} is applied to regular grids of initial conditions in symplectic coordinates $(\tilde{y},\tilde{z})$ (\ref{eq:sym_cords}) over the plane $\phi=0$ for the magnetic fields given by (\ref{eq:fields}). The resolution of the grid is $160\times 160$ initial conditions taken in each sample. Counting the initial conditions that are detected by the method allows to approximately bound from below the area (which in symplectic coordinates represents the toroidal flux) not occupied by tori transverse to the chosen direction. 

Note that areas are the same when computed in symplectic coordinates $(\tilde{y},\tilde{z})$ or in $(\psi,\vartheta)$, because $d\tilde{y}\wedge d\tilde{z} = d\psi \wedge d\vartheta$. Because of this, the area $S$ of nonexistence in the plane $\phi=0$ can be estimated by counting the number of initial conditions detected by Theorem \ref{thm:gen} on a regular grid over $(\tilde{y},\tilde{z},\phi=0)$. In other words, if  $\mathcal{S}$ is the set of points detected by Theorem \ref{thm:gen} on a $N\times N$ regular grid over $[\Tilde{y}_0 - L,\Tilde{y}_0 + L] \times [\Tilde{z}_0 - L, \Tilde{z}_0 + L]$, the area $S$ is approximated by
\begin{equation}
    \label{eq:area}
    S \sim \frac{4L^2}{N^2}\sum_{i=1}^N \sum_{j=1}^N 1_{\{(\tilde{y}_i,\Tilde{z}_j)\in \mathcal{S}\}}
\end{equation}

Also, Theorem \ref{thm:sym} is applied to the same magnetic fields, however on a different set of initial conditions. As the method requires orbits starting from stellarator-symmetric lines, the initial conditions are taken uniformly in $\sqrt{2\psi/B_0}$ along the two semi-lines $\theta=0,\pi$ (where $\tilde{z}=0$) on the $\phi=0$ plane. In particular, it uses a regular partition of $200$ points of the interval $[-1,1]$ in the $\tilde{y}$-axis. We could also have taken initial conditions on the other two half-lines ($\theta=0, \pi$ on $\phi=\pi$), but for the choice of signs of $\eps_{mn}$ that we use, we believe that $\phi=0,\,\theta=0$ is ``dominant'' in the sense that all the primary island chains have an elliptic point on it, and this tends to maximise the set of eliminated trajectories.

The figures that display the results of Theorem \ref{thm:sym} are followed by Poincar\'e sections produced from the iteration of the selected initial points. If any point is detected for nonexistence then so is its whole trajectory; thus even though Theorem \ref{thm:sym} is restricted to symmetric initial conditions, it has implications for a much larger set.	However, estimating the areas occupied by the detected points from the results of this formulation is a more challenging problem, which we hope to address in the future.

To cater for the possibility that the termination condition is never reached on a trajectory, we choose a timeout $t_{f}$. If the timeout is reached then the status of the chosen initial condition is undecided. This, of course, should include all initial conditions that are on invariant tori of the given class, but may include others for which more time would be required to detect the nonexistence. Depending on the implementation, the timeout values might not indicate how long were the trajectories. Thus, in the figures we also display the average of the last computed value of $\phi$ divided by $2\pi$, i.e., the average number of toroidal laps.

For a trajectory, we denote by $t_*$ the time at which non-existence was detected or $t_f$ if it was not detected. As a measure of non-existence of tori of a given class, figures show in hues the relative time of detection, using the ratio $q = t_*/t_{f}$ of time of detection to timeout.

In all the examples throughout this section and elsewhere, we take the following values and function for the vector potential (\ref{eq:potential})
\begin{align}
\label{eq:standard_values}
\begin{split}
\begin{aligned}
w_1& = 1/4,\\
w_2& = 1,
\end{aligned}\qquad
\begin{aligned}
B_0&=1,\\
R_0&=2,
\end{aligned}\qquad
\begin{aligned}
\zeta_{mn}&= 0,\\
f(\psi)&= \psi-R_0^2/B_0.
\end{aligned}
\end{split}
\end{align}
As previously mentioned, the results in all the forthcoming figures are presented over the poloidal plane $\phi=0$. Unless stated otherwise, the timeout used is $t_f=200$, which amounts to $\sim32$ laps around the $z$-axis.

\subsection{Example 1: An integrable case}
The first example considered corresponds to the magnetic field derived from (\ref{eq:potential}) for the resonance $2/1$ (i.e., $(m,n)=(2,1)$). That is,  
\begin{equation}
\label{A_ex1}
A_\phi = -\left[\psi/4 + \psi^2 + \eps\psi (\psi - 4) \cos(2\vartheta-\phi)\right].
\end{equation}
As explained in Subsection \ref{subsec:integrable}, such a field is integrable, lying on surfaces of constant $\Psi=-\,\psi-2A_\phi$.

\begin{figure}[ht!]
 \centering  
    \includegraphics[height=7.3cm, trim={0 0 0 0},clip]{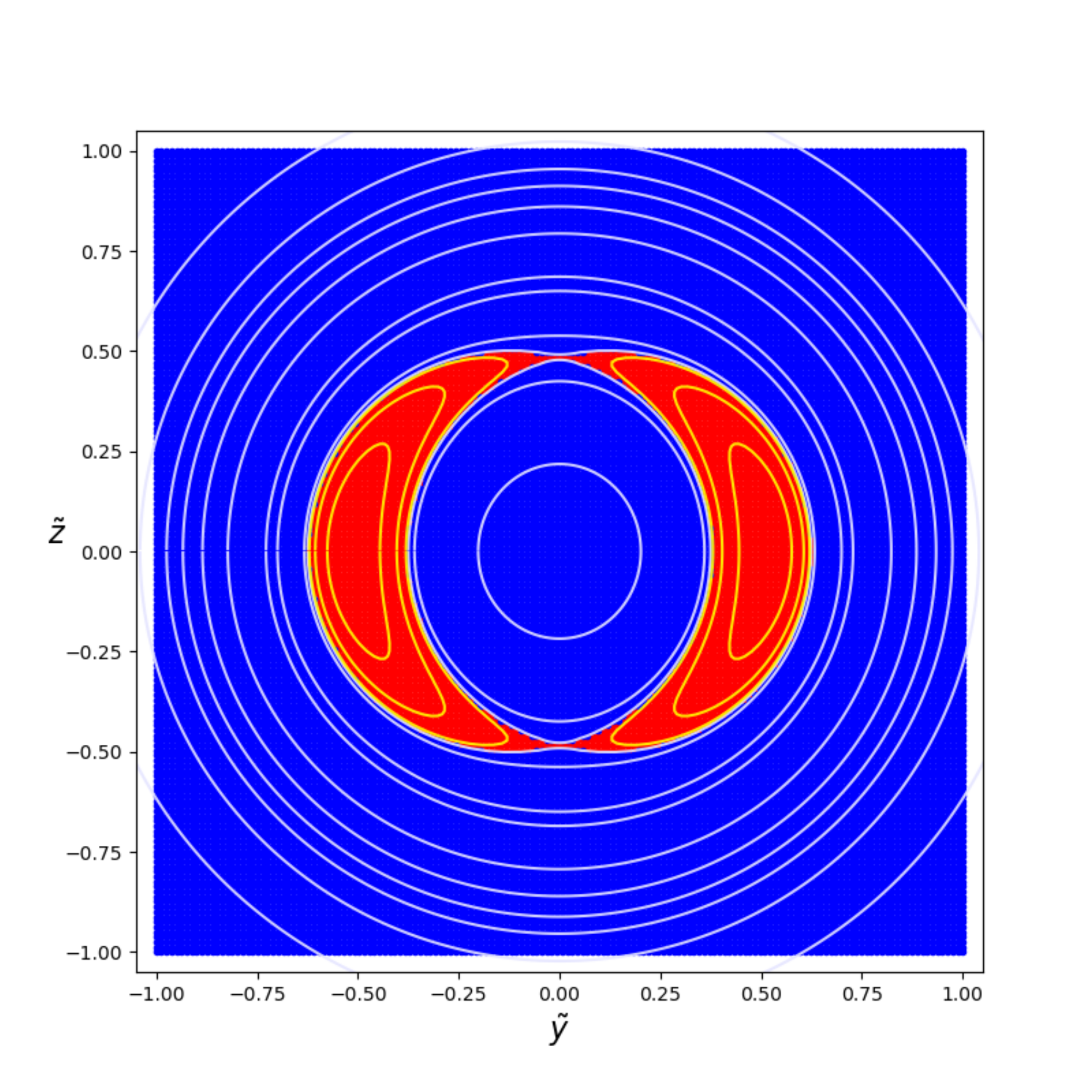}
    \includegraphics[height=7.3cm, trim={0 0 0 0},clip]{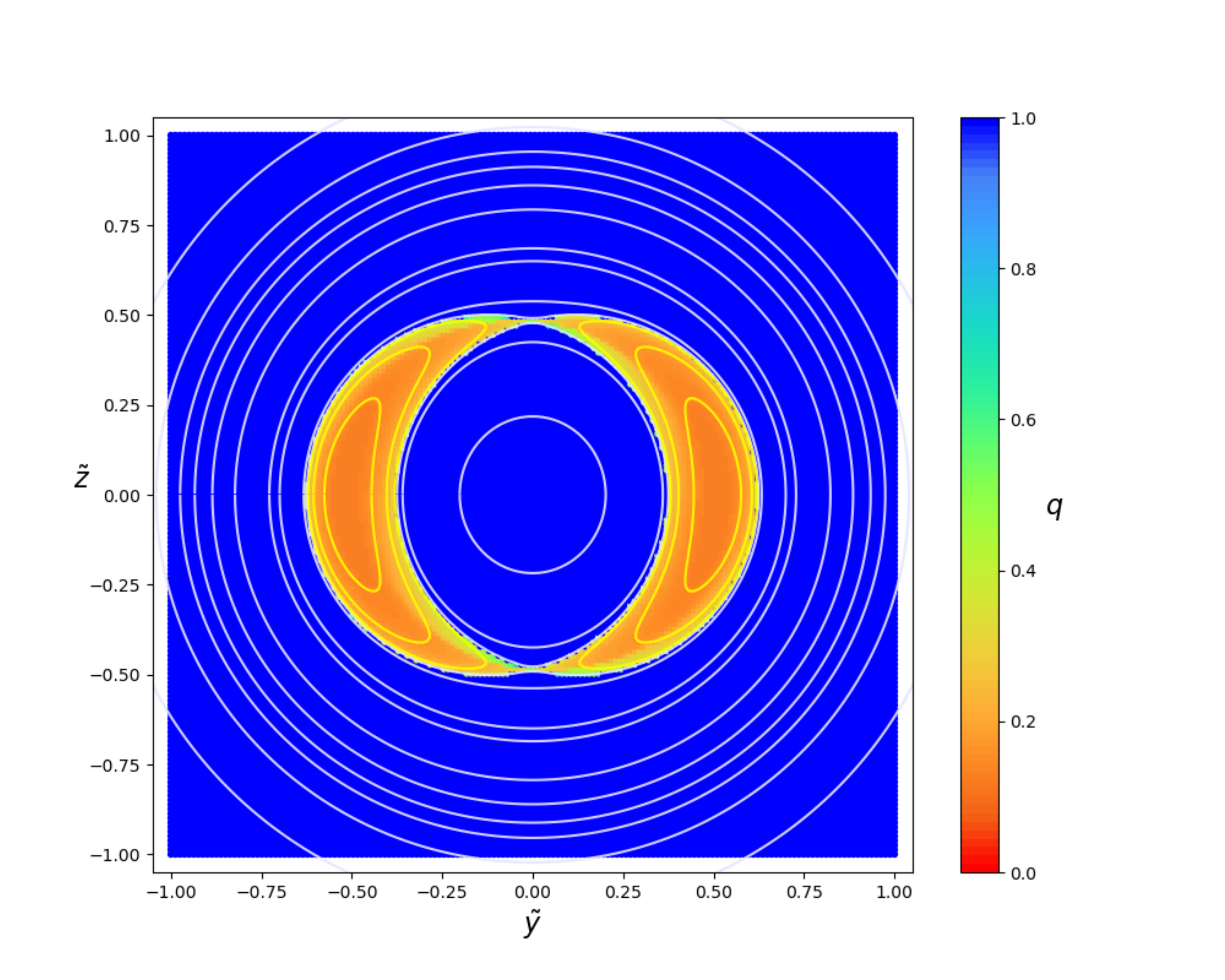}
   \caption{Converse KAM results using Theorem \ref{thm:gen} for Example 1 with $\eps=0.004$ in symplectic coordinates (\ref{eq:sym_cords}). On the left, red = nonexistence, blue = undetermined. On the right, hues vary from fast detection (red) to no detection at all (blue) within timeout. The white curves are level sets of the invariant $\Psi$.}
   \label{cKAM_gen_21}
\end{figure}

The results from the general formulation of Theorem \ref{thm:gen} applied to this case are shown in Figure \ref{cKAM_gen_21}.

The choice of $m=2$ allows an analytical expression for the separatrix delimiting the island. It is given by (\ref{eq:tori}) with the limiting value of $\Psi$ from (\ref{eq:island}). From this, we obtain the width $\Delta\psi$ for the island as a function of $\vartheta$ at given $\phi$:
$$\Delta\psi = \sqrt{\frac{(n/2-w_1-\eps f_0\cos\zeta)^2}{(w_2+\eps f_1\cos\zeta)^2}-\frac{(n/2-w_1-\eps f_0)^2}{(w_2+\eps f_1)(w_2+\eps f_1\cos\zeta)}},$$
with $\zeta = 2\vartheta-n\phi$. 
Recall that $w_2$ corresponds to shear, so if $f_1=0$ we see the familiar behaviour $\Delta\psi \sim 2\sqrt{\eps f_0 (\frac{n}{2}-w_1)}\sin(\zeta/2)/{w_2}$ for small perturbation $\eps$.
The area $S_{\text I}$ of the island(s) can be computed by numerical integration of $\int \Delta\psi\, d\vartheta$. By our choice of coordinates, this is equal to its toroidal flux.

\begin{figure}[ht!]
\centering
    \includegraphics[height=8cm, trim={0 0 0 0},clip]{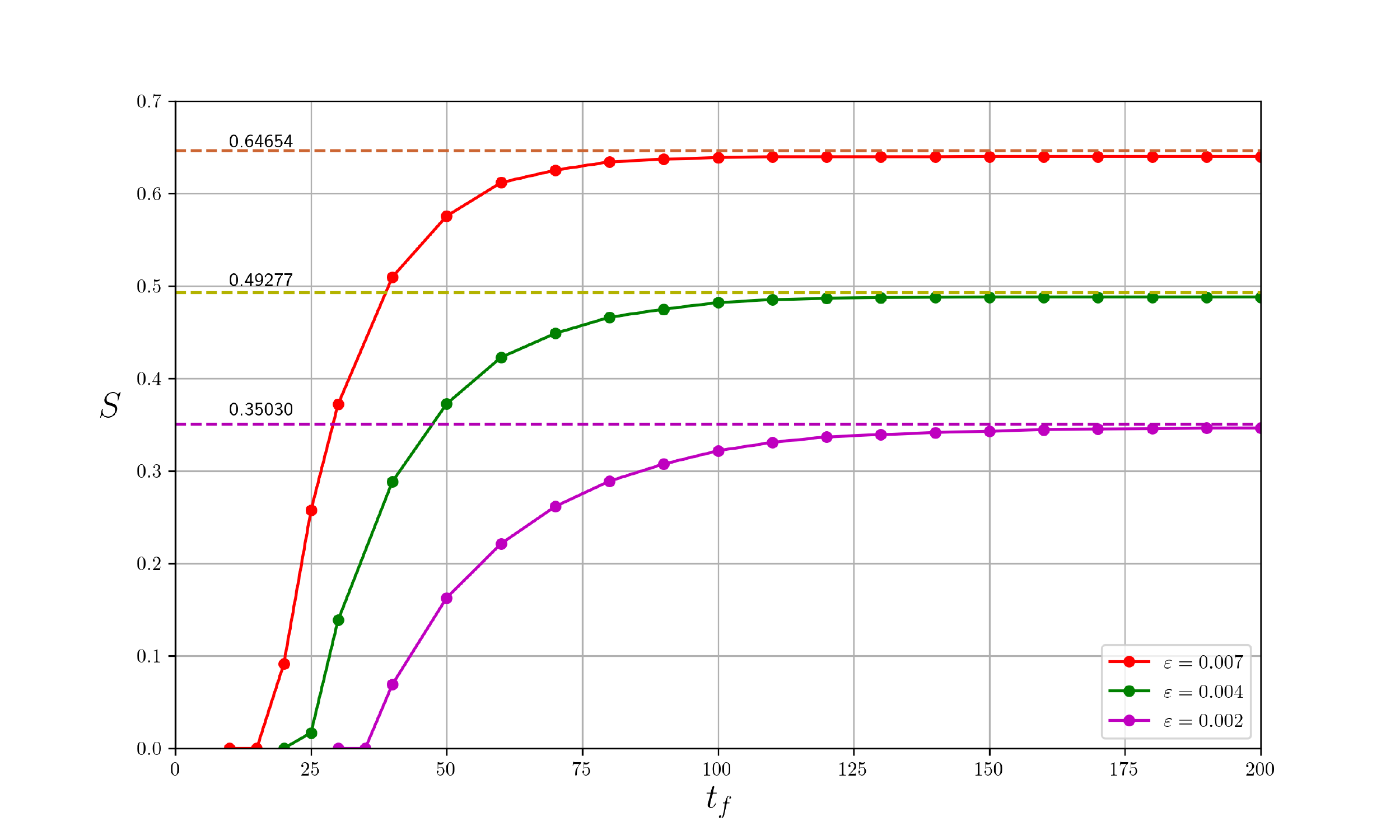}
    \caption{Nonexistence area $S(t_f)$ detected by Theorem \ref{thm:gen} of Converse KAM for different values of $\eps$ for Example 1. The dashed lines represent the corresponding island areas $S_{\text I}$.}
    \label{Fig_area_1res}
\end{figure}

Using (\ref{eq:area}), we calculate the area $S$ of the nonexistence region detected by Theorem \ref{thm:gen}. In Figure \ref{Fig_area_1res}, we see it as a function $S=S(t_f)$ of timeout $t_f$. Recall that timeout units correspond to $t_f/(2\pi)$ laps around the $z$-axis. As expected, the plot shows that the value of the estimated areas increase monotonically with $t_f$ up to a limiting value that agrees with the island area $S_{\text I}$.

Note also that the time of first detection of the island can be predicted:~it is the time for a tangent vector at the centre of the island to make one half of a poloidal revolution (this assumes that the rotation number in the island decreases as distance from the centre increases, else it would be detected earlier). In terms of Greene's residue $R$, which can be written as $\sin^{2}(\alpha/2)$ for eigenvalues $e^{\pm i\alpha}$ of the return map to a poloidal section, we see that for small $R$ (i.e., approximating $R = (\tfrac{\alpha}{2})^2$), the time $t_c$ of first detection should be asymptotically 
\begin{equation}
t_c \sim \tfrac{\pi}{2}T/\sqrt{R}, 
\label{eq:tc}
\end{equation}
where $T$ is the time for one toroidal revolution. 

\begin{figure}[ht!]
\centering
\includegraphics[height=8cm, trim={0 0 0 0},clip]{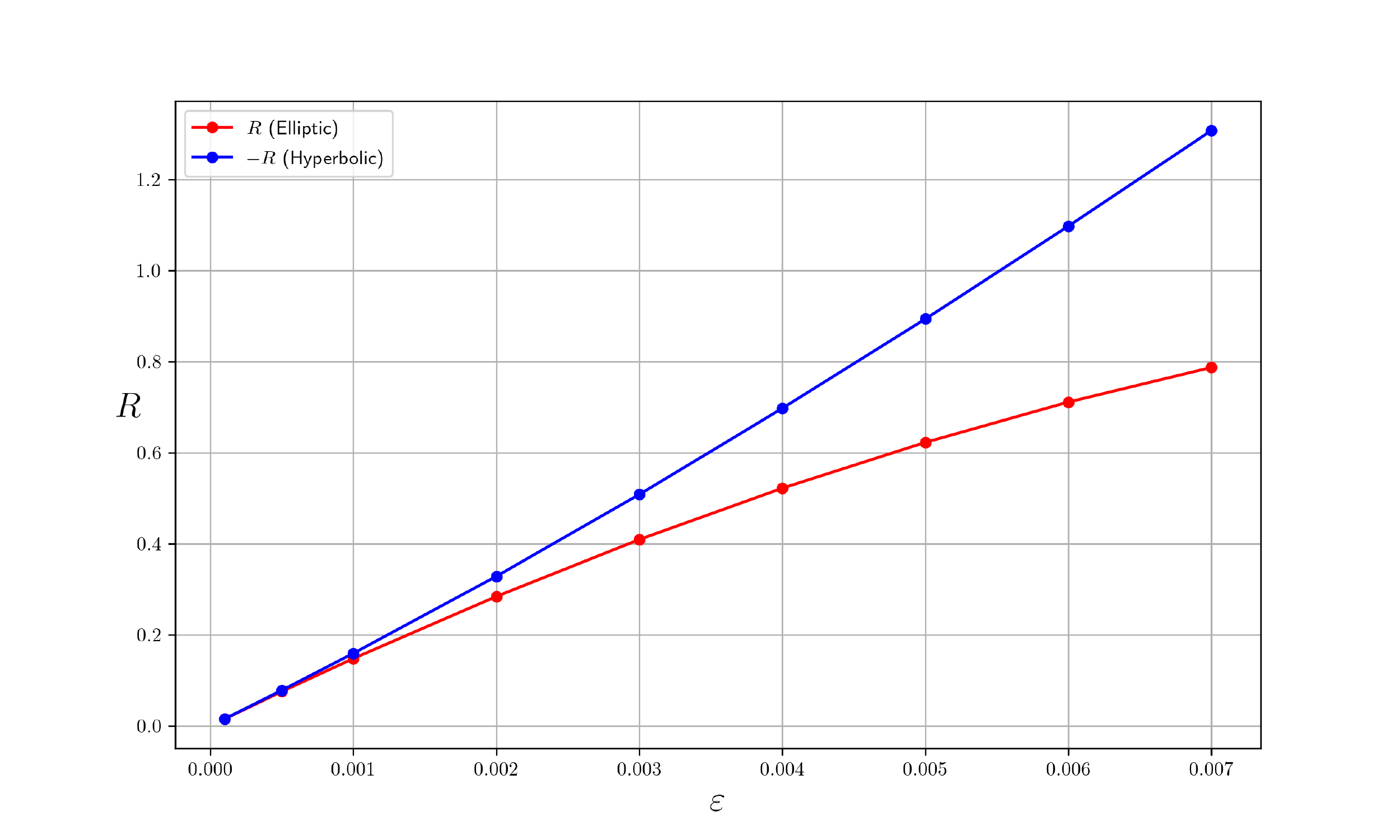} 
\caption{Greene's residue of the island centre (red) and negative residue of the island x-point (blue) for Example 1, as functions of $\eps$.}
\label{fig:residues}
\end{figure}

Figure \ref{fig:residues} shows the numerically computed residue for the centre (and the x-point) of the island as a function of $\eps$. Using the $V$-field, the period of the island centre is $4\pi$, thus $\tfrac{\pi}{2}T \approx 19.74$, and comparing with Figure \ref{Fig_area_1res}, we see that the formula (\ref{eq:tc}) gives a reasonable prediction of the first time of detection of the island. Furthermore, for this example, we see from Figure \ref{Fig_area_1res} that more than half the area of the island has been detected within twice the time of first detection.

\begin{figure}[ht!]
 \centering  
    \includegraphics[height=8cm, trim={0 0 0 0},clip]{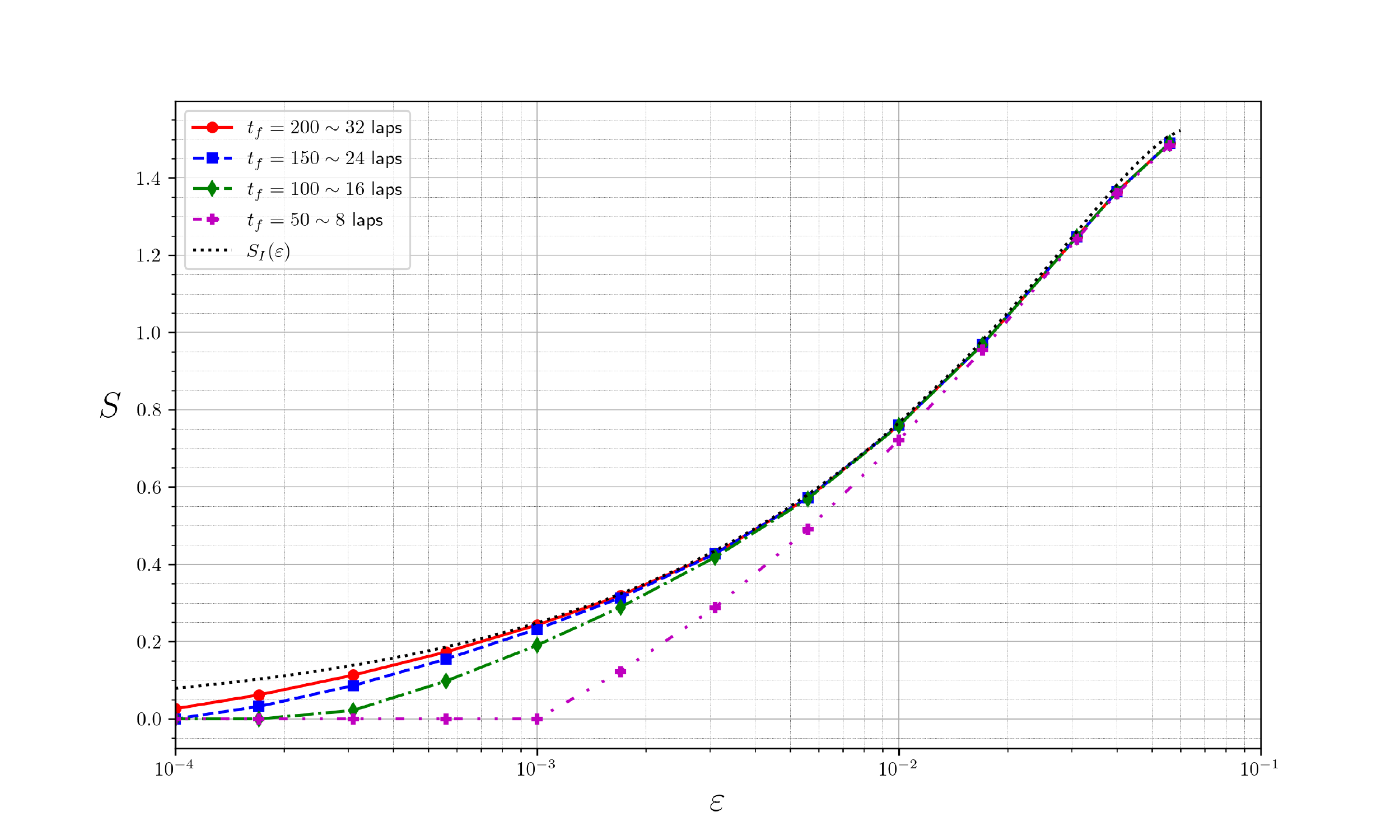} 
   \caption{Nonexistence area $S(\eps)$ detected by Theorem \ref{thm:gen} of Converse KAM for different values of $t_f$ for Example 1, compared with the area $S_{\text I}(\eps)$ of the island.}
   \label{Fig_area_eps_1res}
\end{figure}

The area $S$ of the nonexistence region detected by Theorem \ref{thm:gen} as a function $S=S(\eps)$ of the perturbation parameter $\eps$ is presented in Figure \ref{Fig_area_eps_1res} for different values of timeout $t_f$. It shows that the error of the estimation is reduced for small values of $\eps$ as $t_f$ increases, but ultimately deviates for larger $\eps$. The most likely explanation of this behaviour is that our regular grid does not have enough points in the magnetic island to give a reliable estimation for this region of the parameter.

\begin{figure}[ht!]
\centering  
  \includegraphics[height=7.4cm, trim={0 0 30 0},clip]{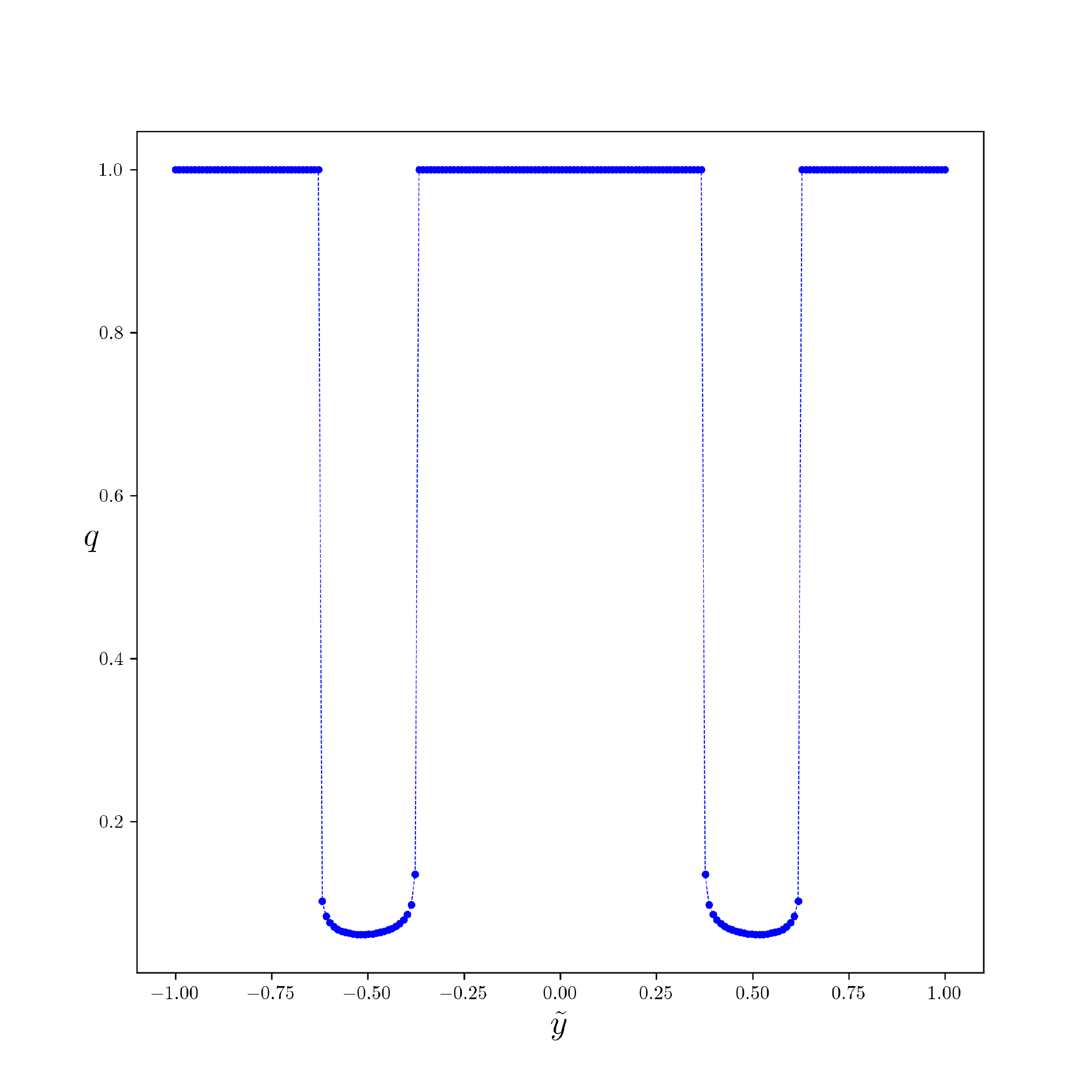}
  \includegraphics[height=7.4cm, trim={0 0 30 0},clip]{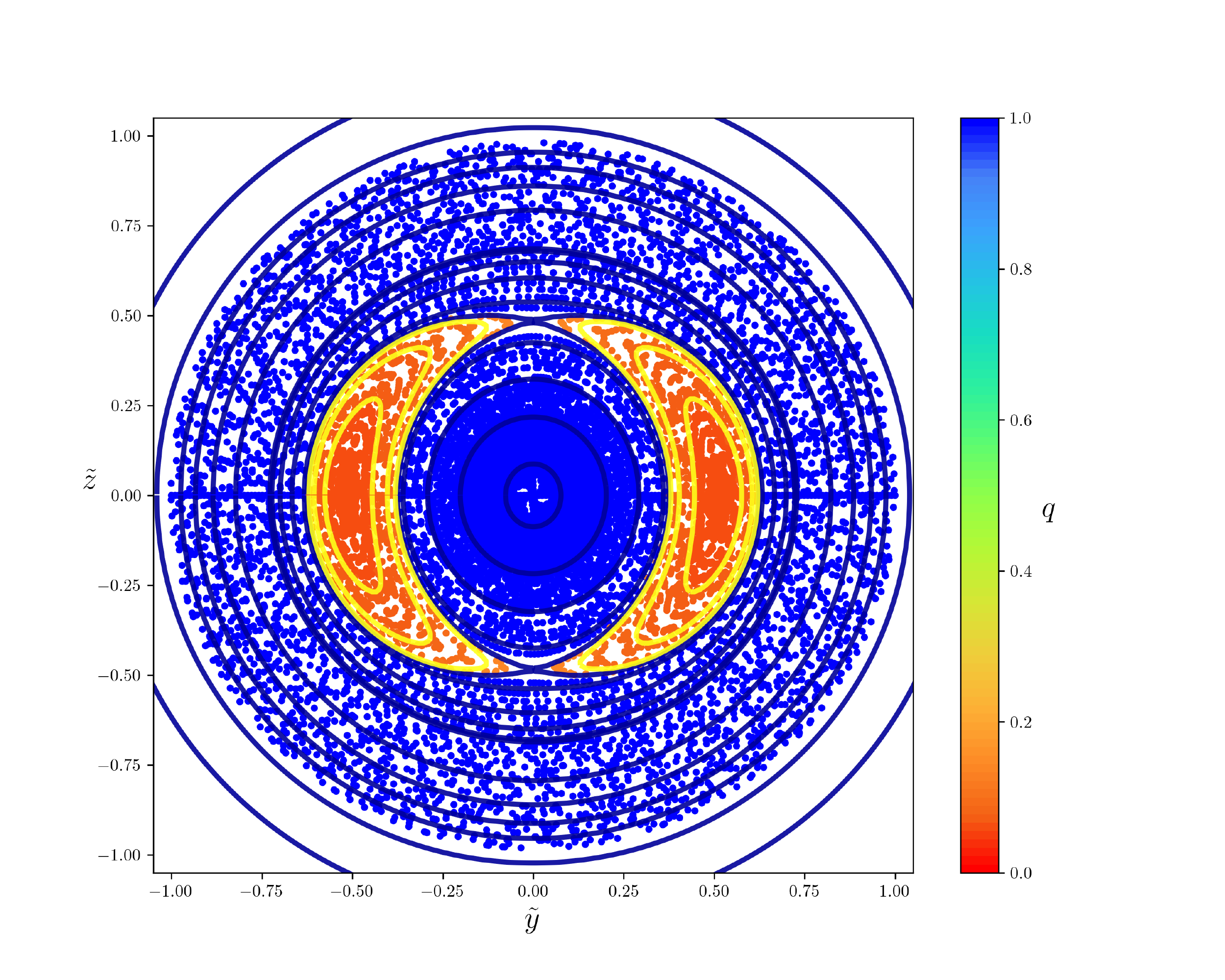}
  \caption{Converse KAM results using Theorem \ref{thm:sym} for Example 1 with $\eps=0.004$, over the symmetrical semi-lines $\vartheta = 0,\pi$ ($\tilde{z}=0$). Hues vary from fast detection (red) to no detection at all (blue) within timeout; level sets of the invariant $\Psi$ are superimposed (right). Relative time of detection as a function of symmetrical initial position (left).}
\label{cKAM_sym_21}
\end{figure}

Figure \ref{cKAM_sym_21} shows the Converse KAM results now using Theorem \ref{thm:sym}. Since this theorem can only be applied on symmetric semi-lines, the results cannot be directly used to estimate the nonexistence area. The left picture shows the relative times of detection in terms of $q=t_*/t_f$ for initial conditions on the partition of the semi-lines. The picture on the right shows the Poincar\'e plot obtained from iteration of the selected initial conditions for the given timeout $t_f$. Compared to the one in Figure \ref{cKAM_gen_21}, it is worth noting that the time of first detection of the island is now half as much, as expected.

\FloatBarrier

\subsection{Non-integrable examples}

Next we consider magnetic fields with more than one helical term, derived from (\ref{eq:potential}). The Poincaré section in all the forthcoming examples displays features of typical near-integrable systems:~tori of different classes and chaotic regions near the hyperbolic saddle of the resonances (magnetic islands). Both formulations, Theorems \ref{thm:gen} and \ref{thm:sym}, yield closely aligned results. Using a radial direction field, they are able to identify and eliminate points (and in fact whole field lines) that do not lie on tori of the original class. They do not distinguish, however, between the ones lying on tori of another class or in chaotic regions. But if required, the use of a suitable foliation centered on the elliptic field lines of an island chain could  differentiate between those two cases.

\subsubsection{Example 2}
\begin{figure}[ht!]
 \centering  
    \includegraphics[height=7.3cm, trim={0 0 0 0},clip]{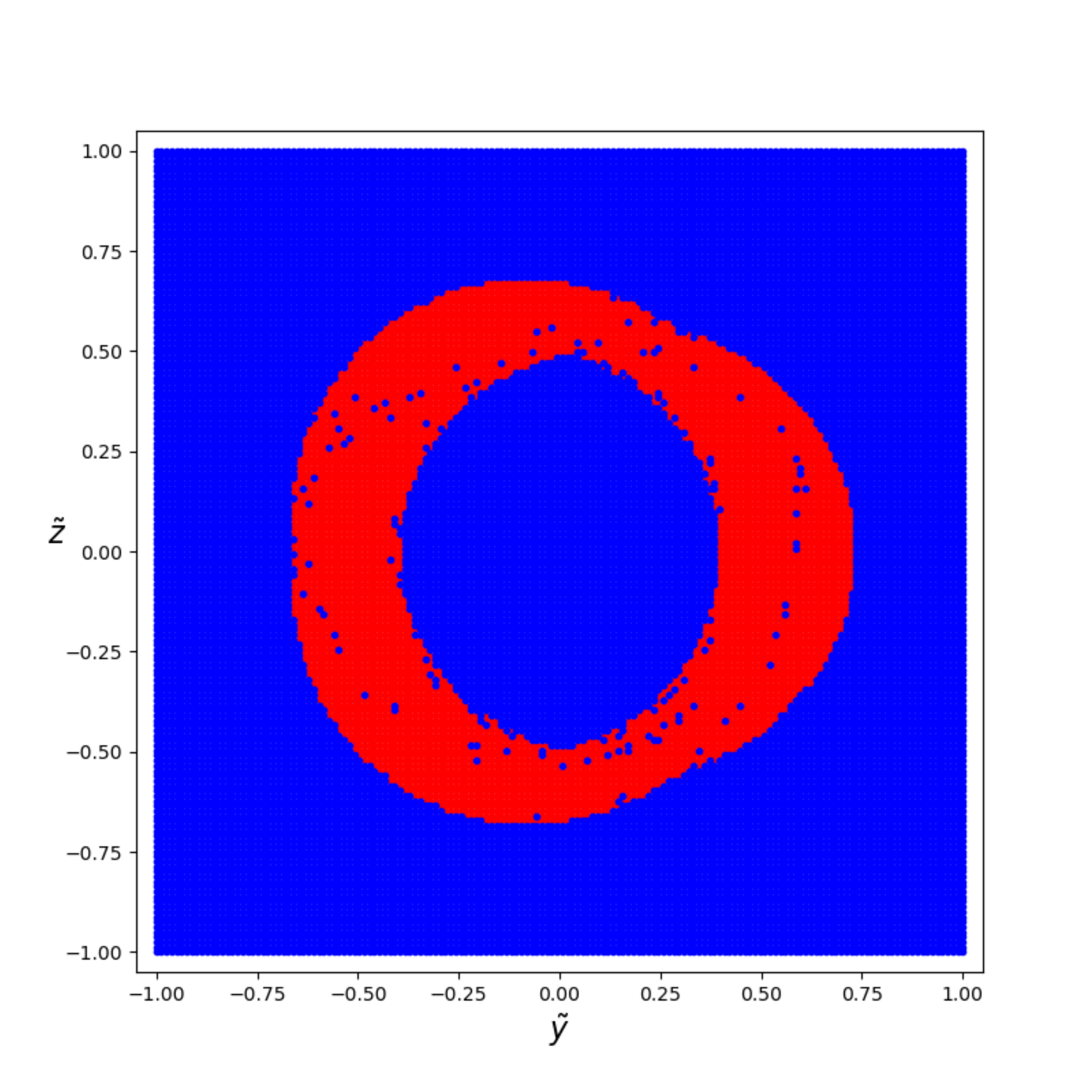}
    \includegraphics[height=7.3cm, trim={0 0 0 0},clip]{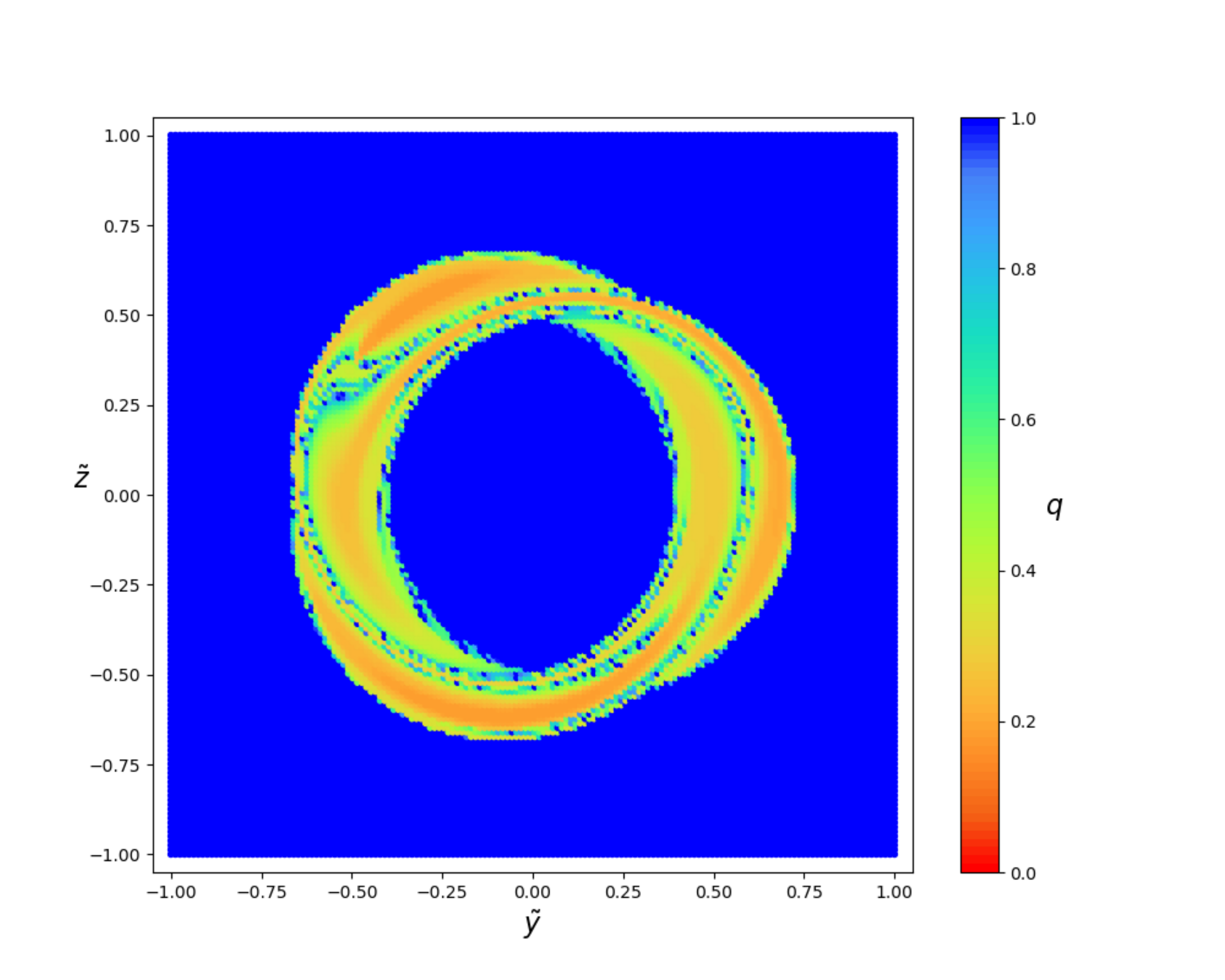}
   \caption{Converse KAM results using Theorem \ref{thm:gen} for Example 2 with $\eps=0.003$ in symplectic coordinates (\ref{eq:sym_cords}). On the left, red = nonexistence, blue = undetermined. On the right, hues vary from fast detection (red) to no detection at all (blue) within timeout.}
   \label{cKAM_gen_21_32}
\end{figure}

\begin{figure}[ht!]
 \centering   
   \includegraphics[height=8cm, trim={0 0 0 1cm},clip]{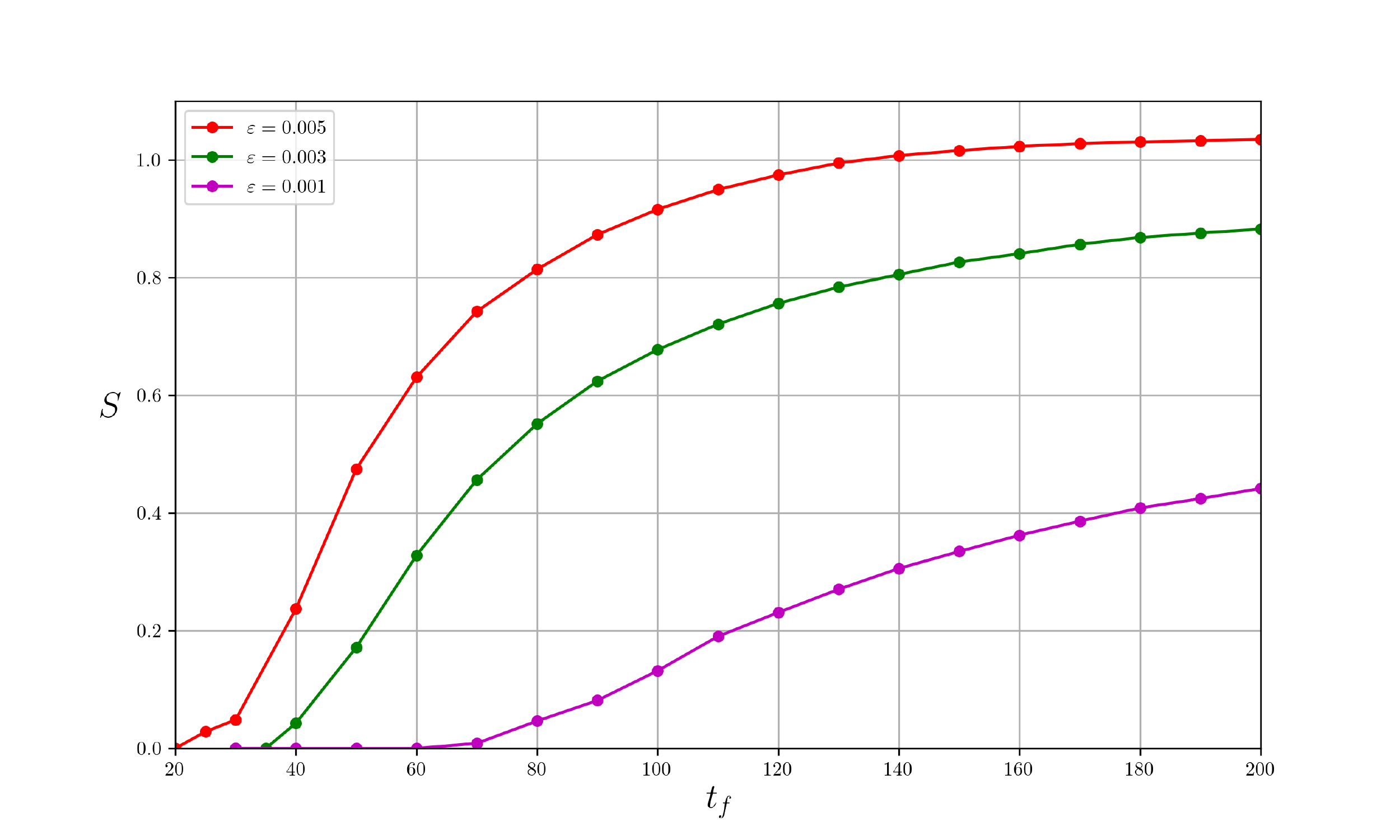}
   \caption{Nonexistence area $S(t_f)$ detected by Theorem \ref{thm:gen} of Converse KAM for different values of $\eps$ for Example 2.}
   \label{Fig_area_2res}
\end{figure}

\begin{figure}[ht!]
 \centering  
   \includegraphics[height=8cm, trim={0 0 0 1cm},clip]{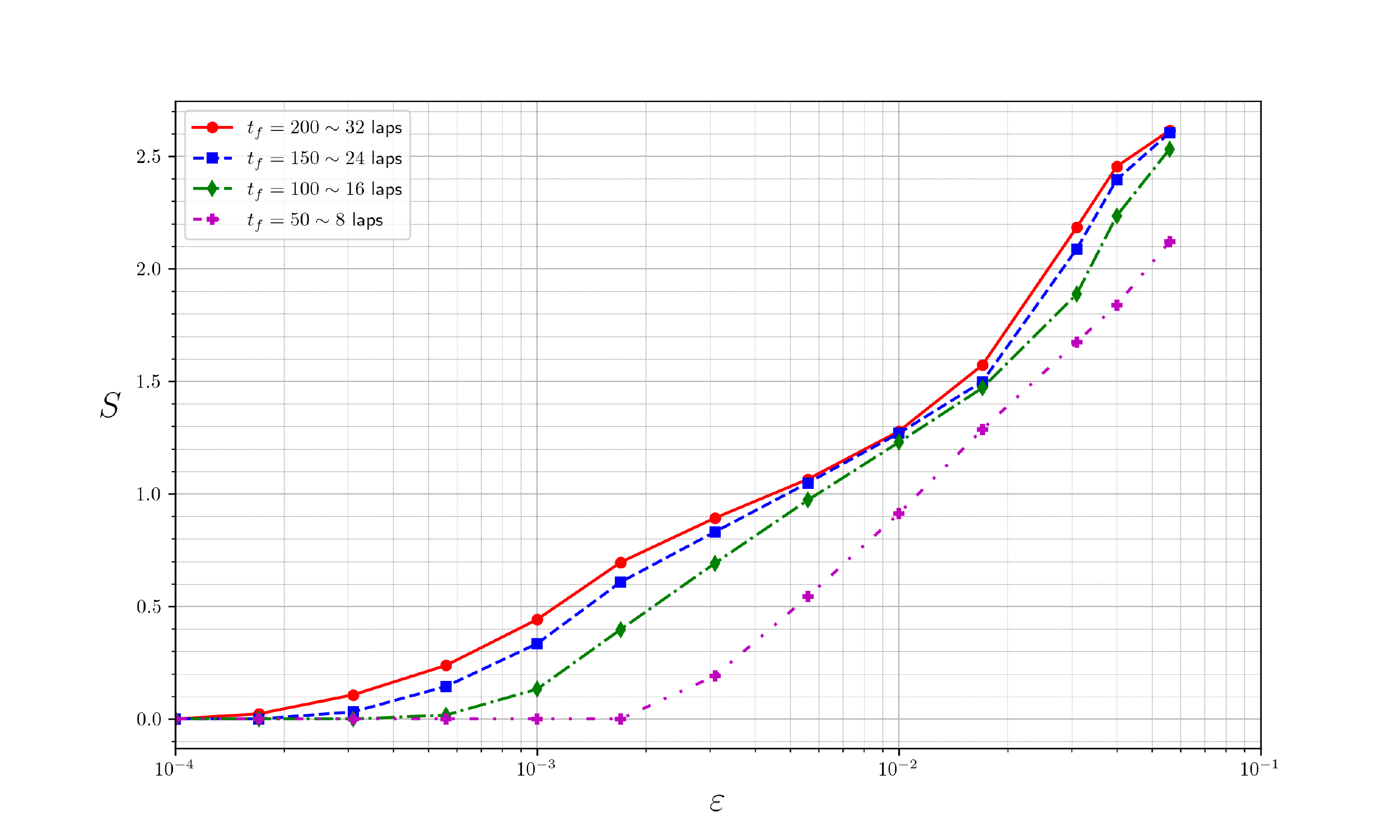}
   \caption{Nonexistence area $S(\eps)$ detected by Theorem \ref{thm:gen} of Converse KAM for different values of $t_f$ for Example 2.}
   \label{Fig_area_eps_Ex2}
\end{figure}

The second example corresponds to the magnetic field derived from (\ref{eq:potential}) for two modes now, namely the resonances $2/1$ and $3/2$ with same perturbation parameter value $\eps_{21} = \eps_{32} = \eps$. That is,
\begin{equation}
\label{A_ex2}
A_\phi = -\left[\psi/4 + \psi^2 + \eps\psi(\psi-4)\left[\cos(2\vartheta-\phi) + \psi^{1/2}\cos(3\vartheta-2\phi)\right]\right].
\end{equation}
Following the same order as in previous example, the Converse KAM results using the formulation of Theorem \ref{thm:gen} are shown in Figures \ref{cKAM_gen_21_32}-\ref{Fig_area_eps_Ex2}.

Figure \ref{cKAM_gen_21_32} shows the detection in symplectic coordinates, using the same color scheme as in Figure \ref{cKAM_gen_21}. As we can see on the right, the different $q$-hues suggest the location of the two magnetic islands corresponding to this example.

Figure \ref{Fig_area_2res} shows the computed area $S=S(t_f)$ of nonexistence from Theorem \ref{thm:gen} for the present example for different values of the perturbation parameter $\eps$. As in Figure \ref{Fig_area_1res}, we see that the estimated areas increase monotonically with $t_{f}$ and seem to be approaching a limit.

Figure \ref{Fig_area_eps_Ex2} shows the estimated area $S=S(\eps)$ of nonexistence from Theorem \ref{thm:gen} now as a function of the perturbation parameter $\eps$ for different values of timeout $t_f$. The behaviour seems to be not as simple as in Figure \ref{Fig_area_eps_1res} for the case of the single resonance. A possible explanation of this may be the interaction between the resonances as they grow with $\eps$.

\begin{figure}[ht!]
\centering  
  \includegraphics[height=7.4cm, trim={0 0 30 0},clip]{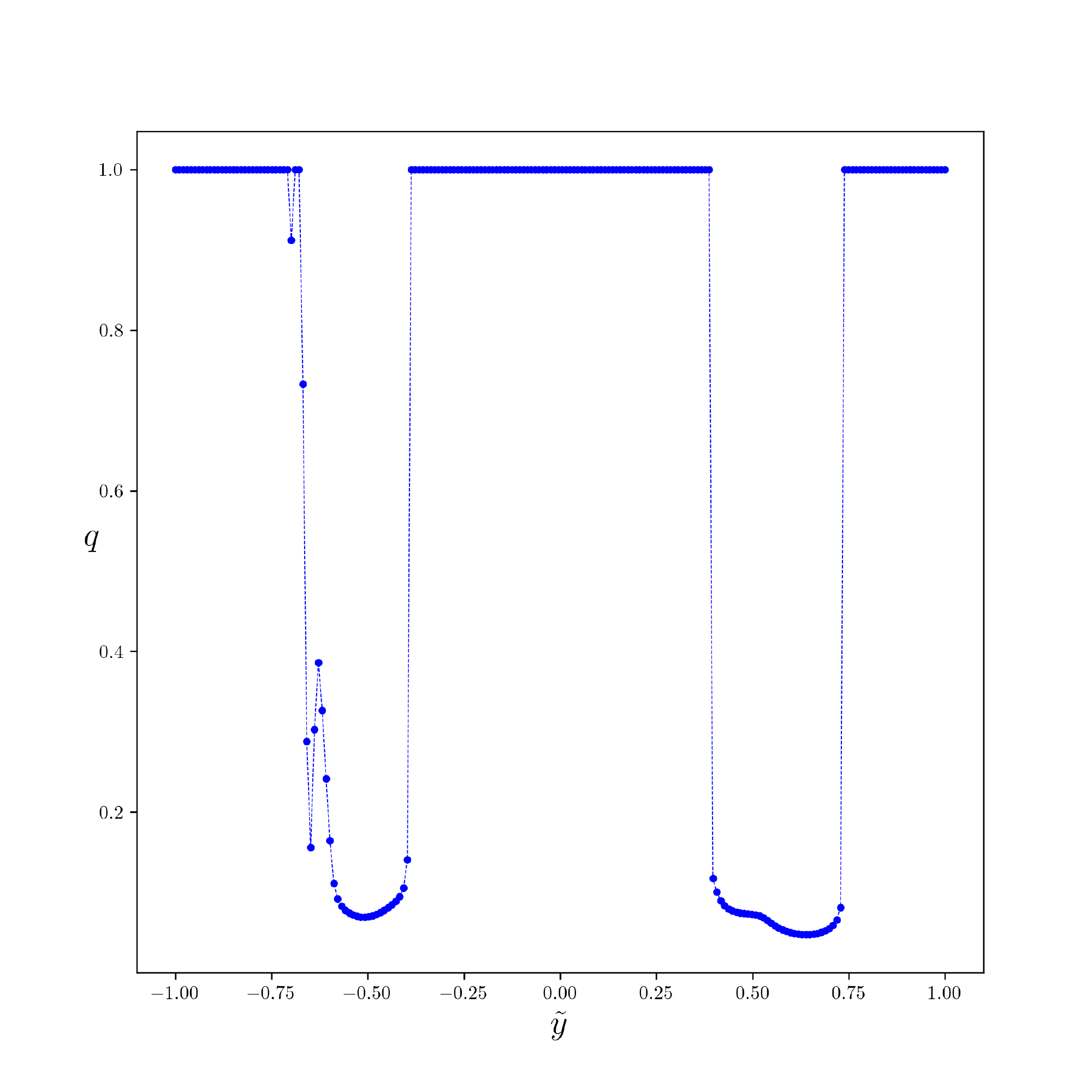}
  \includegraphics[height=7.4cm, trim={0 0 30 0},clip]{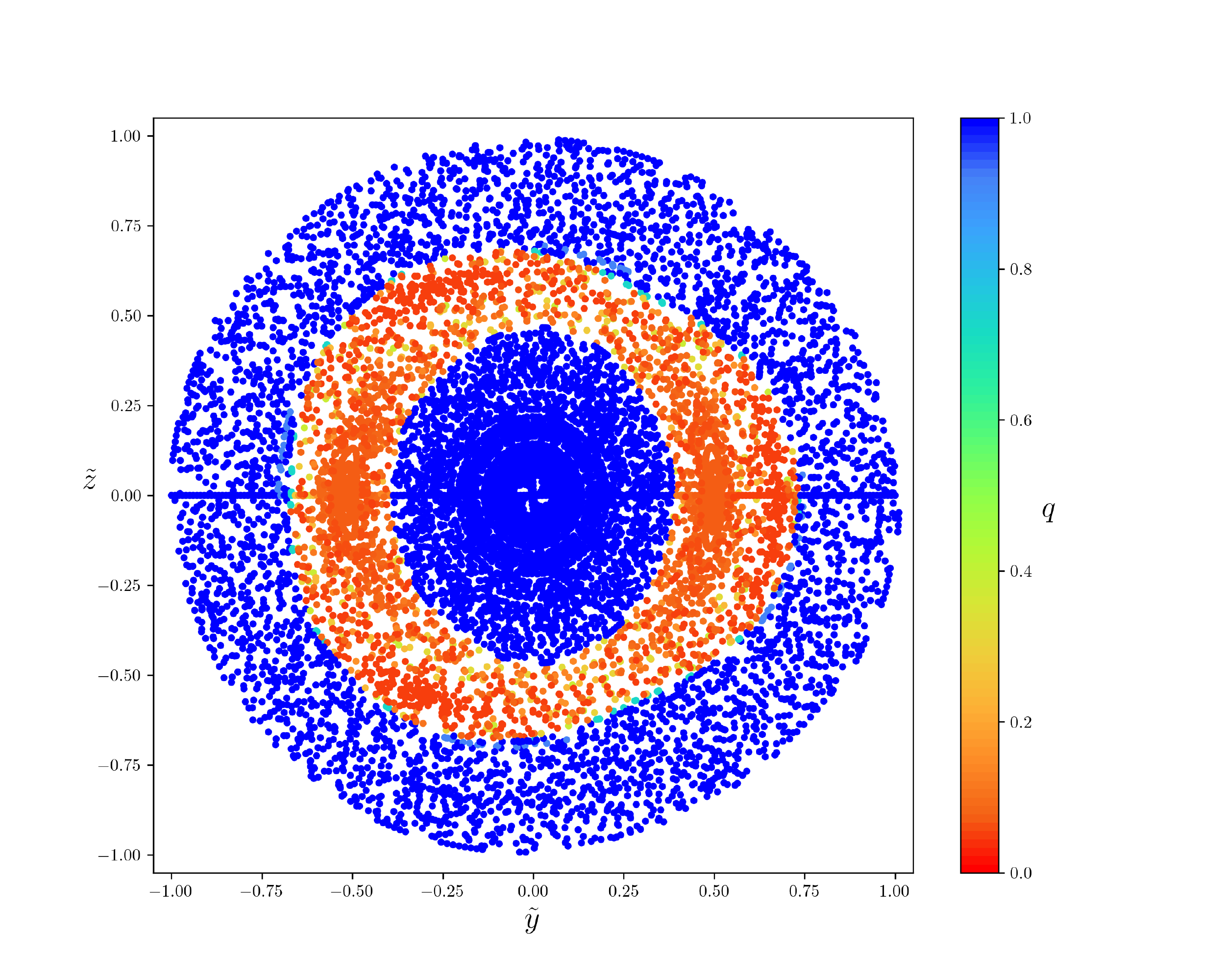} 
  \caption{Converse KAM results using Theorem \ref{thm:sym} for Example 2 with $\eps=0.003$, over the symmetrical semi-lines $\vartheta = 0,\pi$ ($\tilde{z}=0$). Hues vary from fast detection (red) to no detection at all (blue) within timeout (right). Relative time of detection as a function of symmetrical initial position (left).}
\label{cKAM_sym_21_32}
\end{figure}

The Converse KAM results using Theorem \ref{thm:sym} are shown in Figure \ref{cKAM_sym_21_32}. The left plot, compared to the one in Figure \ref{cKAM_sym_21}, shows an asymmetrical distribution of the relative time of detection $q$, which is consistent with the resonances used in this example. 

\FloatBarrier

\subsubsection{Example 3}
\begin{figure}[ht!]
 \centering 
    \includegraphics[height=7.3cm, trim={0 0 0 0},clip]{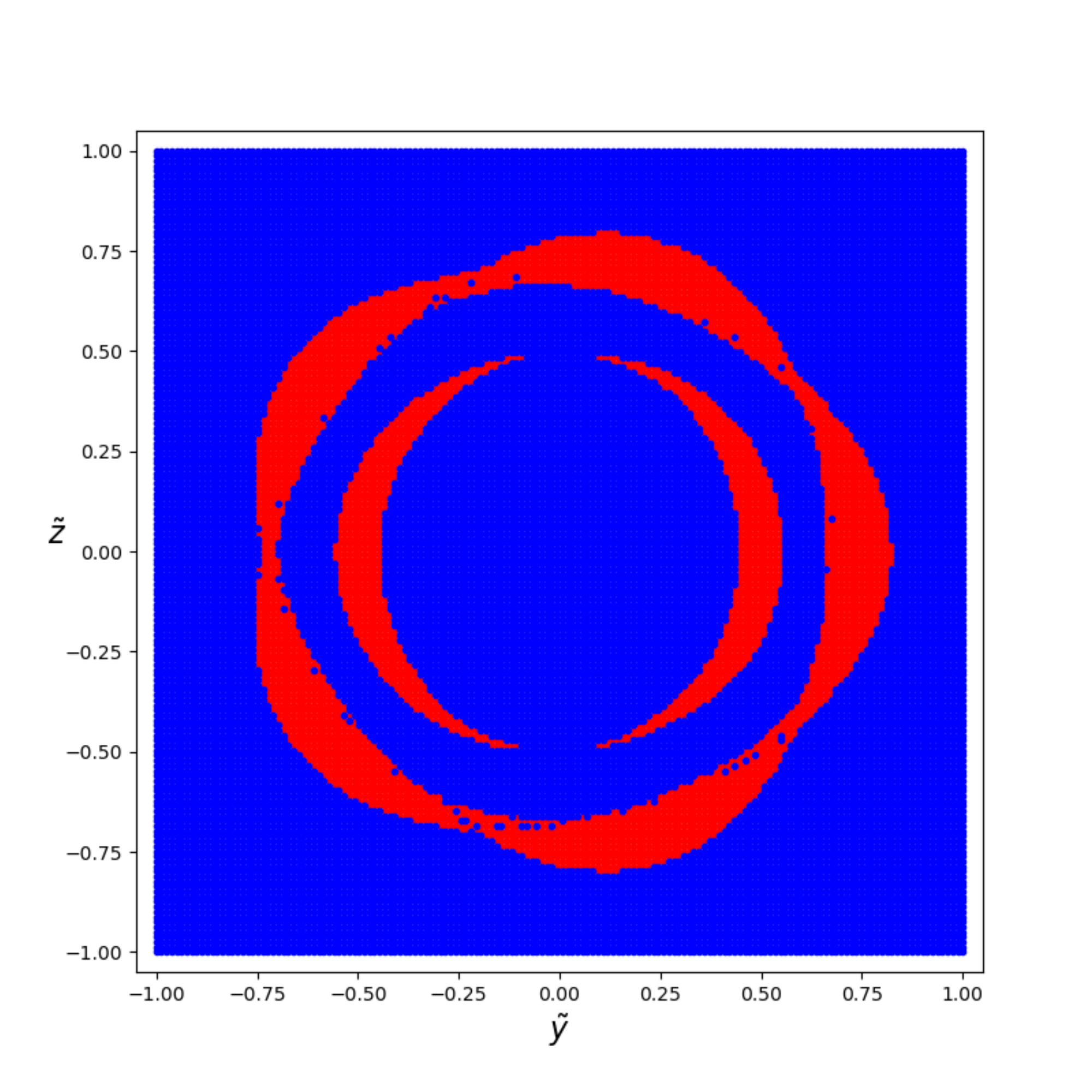}
    \includegraphics[height=7.3cm, trim={0 0 0 0},clip]{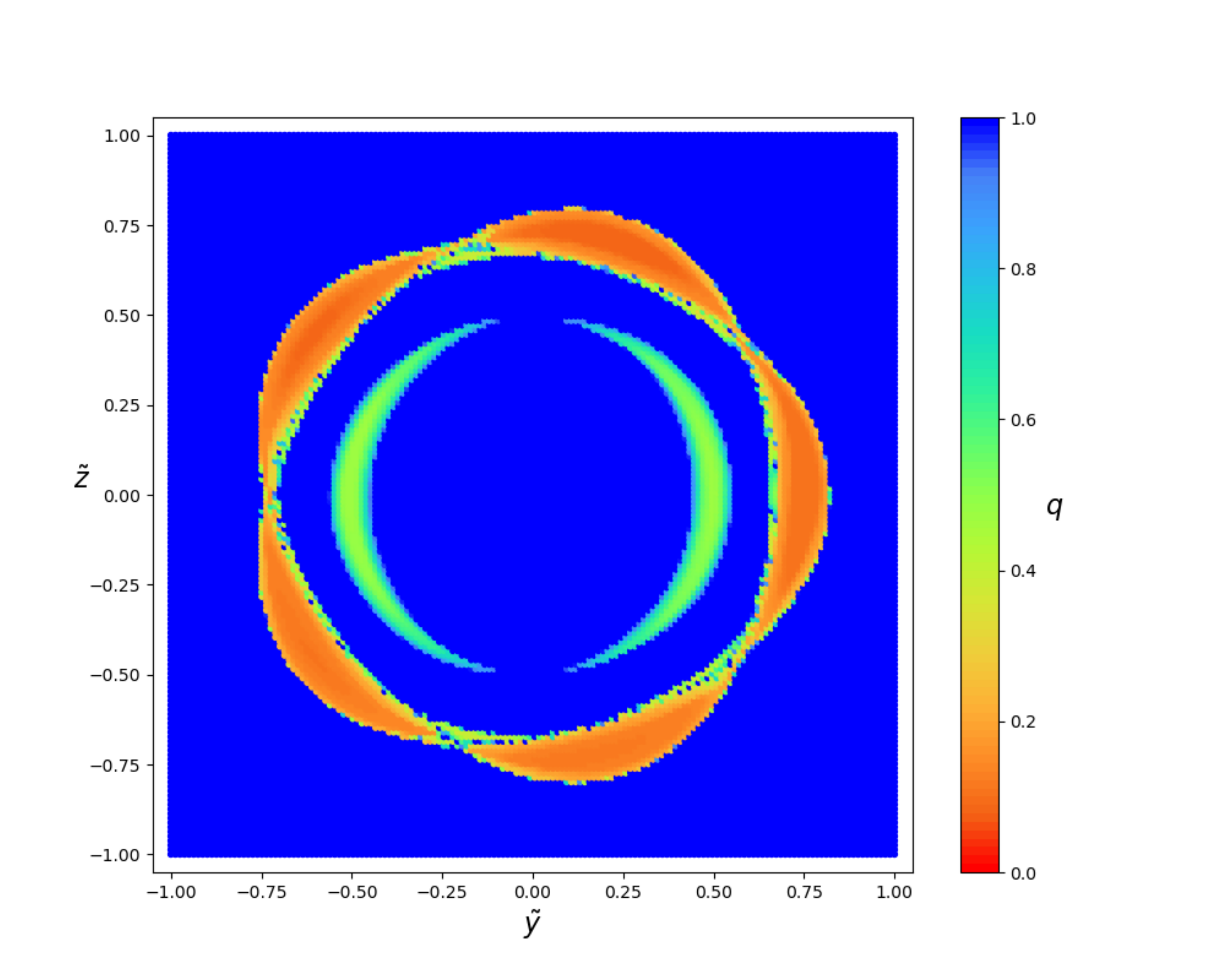} 
   \caption{Converse KAM results using Theorem \ref{thm:gen} for Example 3 with $\eps=0.01$ in symplectic coordinates (\ref{eq:sym_cords}). On the left, red = nonexistence, blue = undetermined. On the right, hues vary from fast detection (red) to no detection at all (blue) within timeout.}
   \label{cKAM_gen_21_54}
\end{figure}

\begin{figure}[ht!]
 \centering    
   \includegraphics[height=8cm, trim={0 0 0 0},clip]{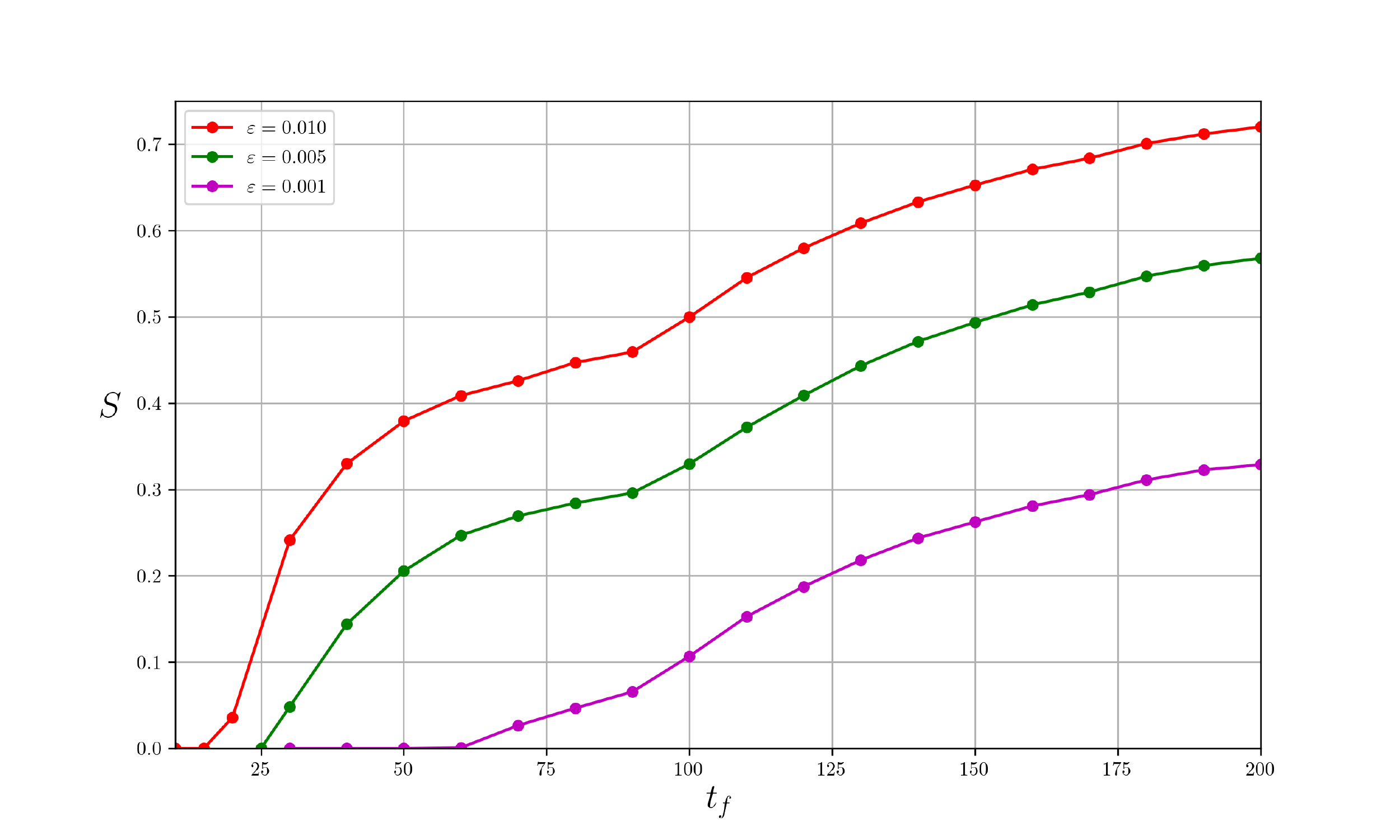}
   \caption{Nonexistence area $S(t_f)$ detected by Theorem \ref{thm:gen} of Converse KAM for different values of $\eps$ for Example 3.}
   \label{Fig_area_21_54}
\end{figure}

\begin{figure}[ht!]
\centering  
   \includegraphics[height=8cm, trim={0 0 0 0},clip]{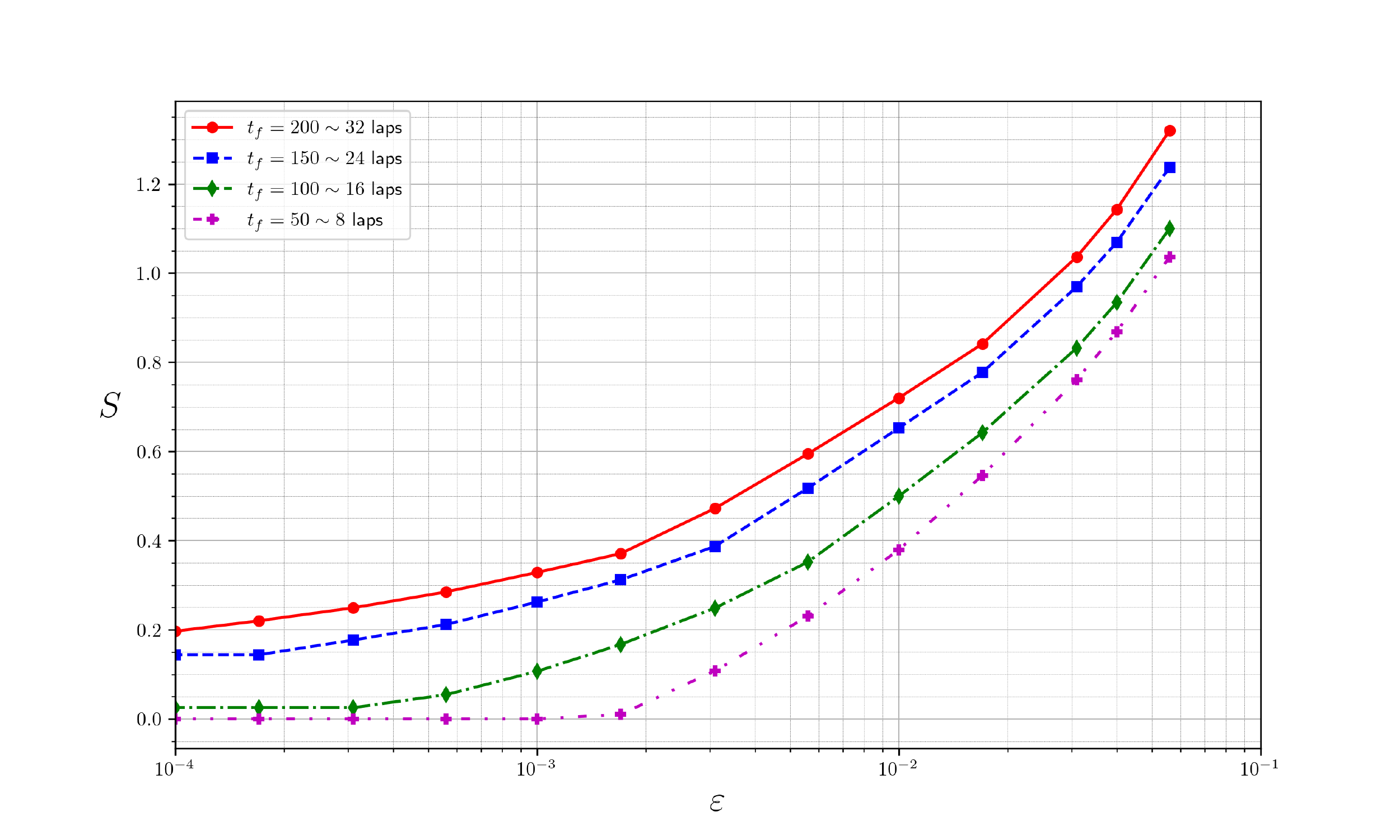}
    \caption{Nonexistence area $S(\eps)$ detected by Theorem \ref{thm:gen} of Converse KAM for different values of $t_f$ for Example 3.} 
    \label{Fig_area_eps_Ex3}
\end{figure}

The next example corresponds to the magnetic field obtained from (\ref{eq:potential}) for the resonances $2/1$ again and $5/4$ now, with fixed value  $\eps_{21} = 0.001$ and varying value of $\eps_{54}=\eps$. That is,
\begin{equation}
\label{A_ex3}
A_\phi = -\left[\psi/4 + \psi^2 +  \psi (\psi - 4)\left[\eps_{21} \cos(2\vartheta-\phi)  +
\eps\psi^{3/2} \cos(5\vartheta-4\phi)\right]\right]. 
\end{equation}
The Converse KAM results using the formulation of Theorem \ref{thm:gen} are shown in Figures \ref{cKAM_gen_21_54}-\ref{Fig_area_eps_Ex3}. Figure \ref{cKAM_gen_21_54} shows the detection in symplectic coordinates for $\eps=0.01$, Figure \ref{Fig_area_21_54} shows the computed area $S=S(t_f)$ of nonexistence for the present example for different values of the perturbation parameter $\eps$, and Figure \ref{Fig_area_eps_Ex3} shows $S=S(\eps)$ for different values of timeout $t_f$. The results behave as expected, except for the particularity that the islands seem to be detected at relatively different times. This is noticeable in the right plot of Figure \ref{cKAM_gen_21_54}, where the resonance $5/4$ is seen mostly in the orange area, while the $2/1$ has a green hue instead. Larger timeout $t_f$ is required for the method to detect magnetic islands of small amplitude (as quantified by the residue).

\begin{figure}[ht!]
\centering  
  \includegraphics[height=7.4cm, trim={0 0 30 0},clip]{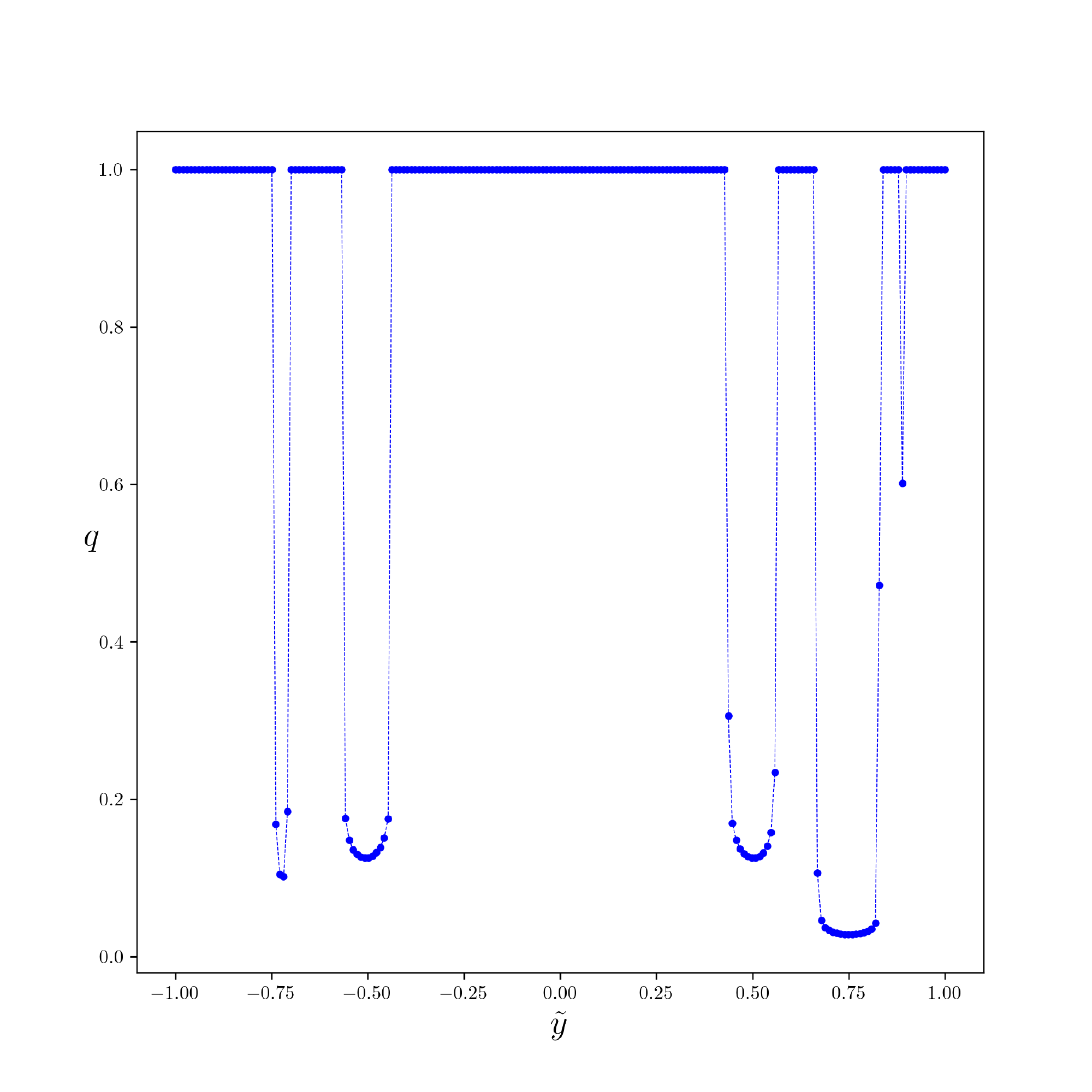}
  \includegraphics[height=7.4cm, trim={0 0 30 0},clip]{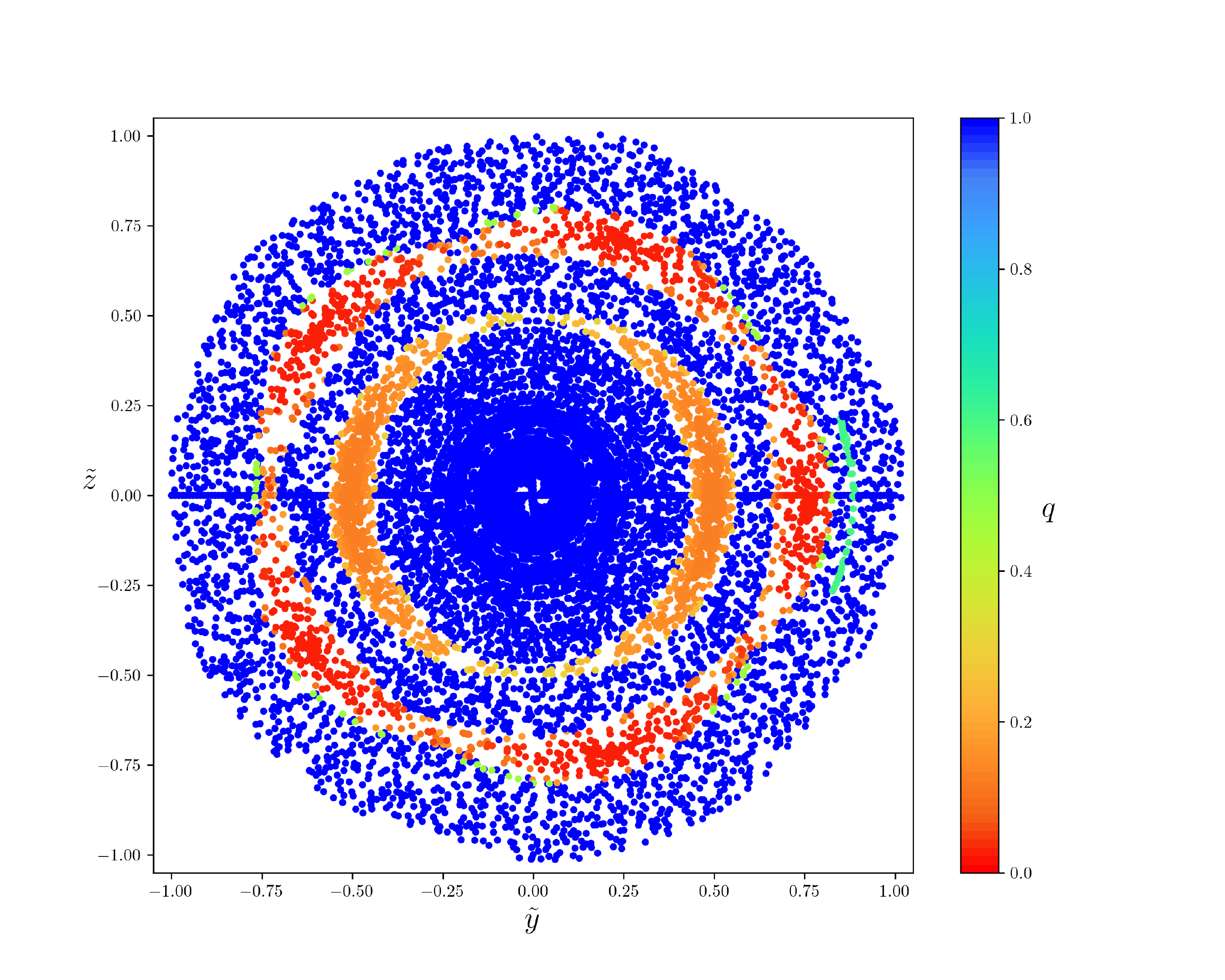} 
  \caption{Converse KAM results using Theorem \ref{thm:sym} for Example 3 with $\eps=0.01$, over the symmetrical semi-lines $\vartheta = 0,\pi$ ($\tilde{z}=0$). Hues vary from fast detection (red) to no detection at all (blue) within timeout (right). Relative time of detection as a function of symmetrical initial position (left).}
\label{cKAM_sym_21_54}
\end{figure}

The corresponding Converse KAM results of Theorem \ref{thm:sym} are shown in Figure \ref{cKAM_sym_21_54}. The results are similar to Figure \ref{cKAM_sym_21_32} for Example 2, besides the different relative detection time $q$ for each magnetic island that we see again.

\FloatBarrier

\section{Conclusions} 
\label{sec:conc}
The paper reports on numerical implementation of the Converse KAM method from \cite{mackay2018finding} on some example magnetic fields. It has demonstrated that the method allows one to identify many of the points that do not belong to any flux surface of a given class. It has been shown to reach decisions in relatively short times on these examples. In an example with an integrable island, it detects a large fraction of the island in a fieldline flow time of order $\tfrac{\pi}{2} T/\sqrt{R}$ (in the symmetric formulation, or twice this for the general formulation), where $T$ is the time for one revolution around the $z$-axis and $R$ is Greene's residue for the island. For fields with stellarator symmetry, it suffices to examine initial conditions just on the symmetry lines.

The method can be used to compute a good lower bound for the toroidal flux that is not on flux surfaces of the desired class.  This is suitable for passing as an objective function to include in optimisation of the design of stellarator fields. One could also compute a lower bound on the {\em volume} not on flux surfaces of given class, by integrating the return time with respect to the toroidal flux over the detected points on a poloidal section.

A crucial component of this method is the selection of a suitable direction field transverse to the class of tori of interest. Our examples have a natural one, which made them a simple test case. For a more general magnetic field, one would need to determine the magnetic axis and a suitable direction field from it, but the freedom to choose the direction field means that the method could in principle be used for fields with bean-shaped cross-section, as in W7X.  Also, one could be interested in survival of tori both around the magnetic axis and in  a major island.  For example, for perturbations of an integrable field with one helical mode, denoting by $\Psi$ the conserved quantity for the integrable case, one could use the direction field $\nabla\Psi$, with respect to some metric.  For some discussion about how to choose the direction field in other contexts, see \cite{duignan2021nonexistence,kallinikos2022regions}.

Although the examples treated here are simple, they already include ones that display the typical mix of islands and chaos.  A next goal is to report on applications of the method to fields produced by the stellarator optimisation code SIMSOPT.  These are designed to be close to integrable, so the aim is to detect and quantify the remaining deviations from integrability, which requires an efficient method, as we believe is ours.

It would be good to implement also the extension of the method called ``killends'', which uses bounds on the slope of invariant tori of given class to extend the region through which they cannot pass \cite{mackay2018finding}. This can eliminate more points without computing more trajectories.  Indeed, it can result in a saving on the total length of trajectories computed, because in essence the trajectories from a grid of points in a transverse section are computed for one revolution, whereas in the general formulation used here, the trajectories from a grid were computed until non-existence detected or timeout, which typically takes many revolutions.

The formulations presented can be extended to guiding-centre motion, which is a 2-parameter family of 3D-systems, parametrised by magnetic moment $\mu$ and energy $E$.  The 3D space is the set of $(x,v) \in \R^3\times \R$ satisfying $\tfrac12 m v^2 + \mu |B(x)| = E$, where $v$ represents parallel velocity. The flux-form $\beta$ is replaced by $e\beta + m d(v b^\flat)$, where $b$ is the unit vector along the magnetic field ($e$ and $m$ are the charge and mass of the particle). One would have to choose appropriate classes of tori for guiding-centre motion. The issues will be somewhat similar to those for the planar circular restricted three-body problem, treated in \cite{kallinikos2022regions}.

In the magnetic field context, a similar method (``phase rotation'') was introduced by White \cite{white2012} (see also Fig 6.8 in \cite{white2013theory}) and applied to guiding-centre motion in a magnetic field, but in our opinion it needs some clarification.  Firstly, it needs stating that the class of tori under consideration are the graphs of functions $P = P_\zeta(\theta)$ in the given coordinate system.  Secondly, it is  stated that the angle $\chi$ for the displacement vector between orbits on different invariant tori cannot rotate by more than $\pi$, but should specify that this means that relative to the vertical, $\chi$ has to remain in $(-\pi,\pi)$.  Our method can be considered to be the limiting case from displacement to tangent vectors, which may be more effective because the displacement vector from a trajectory in an island to a trajectory with a different rotation number might not rotate by more than $\pi$ whereas an infinitesimal one will.
Moreover, our method does not require a Poincar\'e section, and extends to other classes of tori.
White's method has the advantage relative to \cite{mackay1989criterion} that it does not require shear, though it was realised some time ago that at least the 3D case of \cite{mackay1989criterion} did not require shear (leading eventually to \cite{mackay2018finding}).

The method can be applied to  other plasma physics problems too.  For example, for an interface in a stepped pressure equilibrium to support a pressure jump, one needs an invariant torus of the ``pressure-jump'' Hamiltonian.  Some conditions under which none exist were determined by \cite{kaiser1994surface}, but it would be useful to extend them using the Converse KAM method.  This would, for example, shed light on the work of \cite{qu2021non}.  The standard direction field is the relevant one for this problem, so it suffices to use \cite{mackay1989criterion} rather than the current paper. Note that that paper applied the Converse KAM method to another plasma physics problem: the motion of a charged particle in the field of two electrostatic waves.
Further plasma physics examples suggested by a reviewer include time-dependent fields such as those arising from resonances between particles and Alfv\'en modes \cite{white2016} and the quasi-single-helicity states reported in \cite{veranda2019}.

\subsection*{Acknowledgments}
This work was supported by a grant from the Simons Foundation (601970, RSM). We are grateful for the insightful comments and suggestions from J. Meiss, J. Loizu, M. Landreman and E. Paul.

\subsection*{Code availability}
Code for the computations is available on \url{https://github.com/dvmtz-1/cKAM} 
\cite{cKAM2023}.

\begin{appendices}

\section{Some pedagogy}
\label{app:pedagogy}
\subsection{Magnetic flux-form}
It is usual to define a magnetic field in 3D as a divergence-free vector field $B$, but it can equivalently be defined as a closed 2-form $\beta$, where $\beta(\xi,\eta)$ represents the magnetic flux through the infinitesimal parallelogram spanned by the ordered pair of tangent vectors $(\xi,\eta)$ at a point. The relation between the flux-form and the field is
$$\beta = i_B\Omega,$$
where $\Omega$ is the volume-form (non-degenerate 3-form) with respect to which $B$ is divergence-free ($\textrm{div} B = 0$ if and only if $\beta$ is closed: $d\beta = 0$).
The formula says how to obtain $\beta$ from $B$:~$\beta(\xi,\eta) = \Omega(B,\xi,\eta) = B \cdot (\xi \times \eta)$ in vector calculus notation.

Conversely, given a 2-form $\beta$ and a volume-form $\Omega$ in 3D, one can obtain vector field $B$ at any point by noting that the kernel of $\beta$ ($\ker \beta = \{u : i_u\beta=0\}$) is non-zero because $\beta$ is antisymmetric and dimension 3 is not even.  Take any non-zero $u \in \ker \beta$ and extend to a basis $(u,v,w)$; let $B = \frac{\beta(v,w)}{\Omega(u,v,w)} u$.  Then the resulting $B$ does not depend on the choices of $(u,v,w)$ and satisfies $i_B\Omega = \beta$.

Note that representation of magnetic field lines by a closed 2-form $\beta$ does not require the volume-form $\Omega$. The fieldlines are the integral curves of $\ker \beta$.  All the volume-form does is to define the speed at which $B$ goes along them.

\subsection{Vector potential}
It is commonplace that for a 3D vector field $B$, $\textrm{div}\, B = 0$ iff there exists a vector field $A$ such that $B=\textrm{curl}\, A$.  We explain here the related result for 2-forms.

First, we note that the above statement is not quite true for general 3D manifolds.  One should strengthen the definition of a magnetic field from $\div\, B = 0$ to $\int_S B\cdot dS=0$ for every closed surface $S$ (this is often called ``absence of magnetic monopoles'').  The corresponding strengthening of the definition of a magnetic flux form $\beta$ is that $\int_S \beta = 0$ for all closed surfaces $S$.

It follows that there exists a 1-form $\alpha$ such that $\beta = d\alpha$.  Such an $\alpha$ is called a potential for $\beta$.  Given a vector potential $A$ for $B$, a potential for $\beta$ is $\alpha = A^\flat$, where $A^\flat (\xi) = A\cdot \xi$ for all tangents $\xi$.  Conversely, any potential $\alpha$ defines a vector potential $A$.  Just as $A$ is non-unique up to addition of any gradient (even multi-valued), $\alpha$ is non-unique up to addition of any closed 1-form.  The relation $\beta = d\alpha$ is equivalent to $B=\textrm{curl}\, A$, but is simpler because it makes no use of a Riemannian metric.

\subsection{Integrable fields}
We say a magnetic field $B$ is integrable if there is a function $\Psi$ with non-zero derivative almost everywhere such that $B\cdot \nabla \Psi= 0$.  It follows that the surfaces of constant $\Psi$ are invariant under the fieldline flow.

Integrability is nicely addressed at the level of continuous symmetries of potentials for the flux-form.  If there is a vector field $u$ such that the Lie derivative $L_u\alpha = df$ for some function $\zeta$ then using $L_u = i_ud+di_u$ on differential forms one obtains 
$$i_u\beta = d\Psi,$$
with $\Psi=f - i_u\alpha$.  This says that $i_ui_B\Omega = d\Psi$, so in particular $i_Bd\Psi=0$ by antisymmetry of $\Omega$.  In vector calculus these relations are $B\times u = \nabla \Psi$ and $B\cdot \nabla \Psi = 0$.  If $u$ is independent of $B$ almost everywhere then $d\Psi \ne 0$ almost everywhere.

\section{Curvilinear coordinates \& components}
\label{app:curv}

Magnetic fields are often presented in adapted coordinate systems, in particular to make an elliptic closed fieldline into an origin for a toroidal system of coordinates with a radial coordinate and two angle coordinates, and level sets of the resulting radial coordinate to be approximate flux surfaces.  Thus it is important to be able to manipulate the components of the field in such a coordinate system.  Furthermore, for the application of the Converse KAM method it is natural to use the adapted coordinate system to define the principal class of tori of interest to be the tori that are transverse to each of the curves of constant angle coordinates (though there is generally some freedom in choice of origin of the angle coordinates as a function of radial coordinate, so there is not a unique prescription and one might prefer gradient curves of the radial coordinate with respect to some metric).

Thus, it is important to be able to apply the Converse KAM method in a general coordinate system. This requires an understanding of components of vector fields and differential forms in a general coordinate system.
We claim that it is simpler to use covariant and contravariant components than physical components.  These are described in various places with particular reference to plasma physics, e.g.~\cite[Appendix G2]{balescu1988transport}, \cite[Chapter 1]{white2013theory}, but we feel it helpful to give our own perspective here.

Let $x^i, i=1,\ldots d,$ be coordinates on an open subset $U$ of a $d$-dimensional manifold $M$ ($d=3$ in our case), i.e.~differentiable functions $x^i : U \to \R$ whose derivatives $dx^i$ are linearly independent at each point of $U$.  A vector at a point is the velocity of a smooth parametrised curve through the point.  A {\em covector} at a point of $U$ is a linear map from vectors to $\R$.  A 1-form is a smooth choice of covectors on $U$.  Then any 1-form $\alpha$ on $U$ can be written uniquely as $\alpha_i dx^i$ (with summation convention) for $d$ functions $\alpha_i : U \to \R$ called the components of $\alpha$ (they are often called ``covariant components'').

The coordinate functions $x^i$ also induce a special set of vector fields $\partial_i$ on $U$, defined by thinking of them as differential operators on smooth functions $f:U\to \R$ defined by $\partial_i f = \frac{\partial f}{
\partial x^i}$ keeping the other $x^j$ fixed (this is the rate of change of $f(x(t))$ along a curve with $\dot{x}^i = 1$ and $\dot{x}^j = 0$ for $j\ne i$).  The $\partial_i$ form a basis at each point of $U$, called covariant basis, so any vector field $B$ on $U$ can be written uniquely as $B^i \partial_i$ for $d$ functions $B^i : U \to \R$, called the components of $B$ (often called ``contravariant components'').  In particular, the equations for motion along a vector field $B$ in coordinate system $(x^i)$ are just $\frac{dx^i}{dt} = B^i$.

The 1-forms $dx^i$ and vector fields $\partial_j$ are related by $dx^i (\partial_j) = \delta^i_j$, the Kronecker-delta.  Equivalently, $\partial_j x^i = \delta_j^i$.

Complications arise when the manifold comes with a Riemannian metric, because that allows firstly to introduce the idea of unit vectors and secondly to convert between vector fields and 1-forms.

A Riemannian metric is a smooth choice of inner product $\la,\ra$ on each tangent space. In coordinate system $(x^i)$, it can be written as 
$\la u, v \ra = g_{ij}u^i v^j$ for a symmetric set of functions $g_{ij}$ forming a positive-definite matrix at each point.
For a vector $v$ at a point, $\|v\|=\sqrt{\la v,v \ra}$ is called its length.  

In particular, $\|\partial_i\| = \sqrt{g_{ii}}$. This leads to consider the normalised basis $\partial_i /\sqrt{g_{ii}}$ for vectors at a point and hence the {\em physical components} of a vector $B$, namely $\tilde{B}_i = \sqrt{g_{ii}} B^i$ (no summation).  Most of the physics literature uses physical components, yet they are somewhat artificial and lead to extra factors in many formulae, e.g.~for motion along a vector field, and grad, div and curl.

Associated to a Riemannian metric is a natural bijection between vectors and covectors.  Given a vector $v$ at a point,  define the covector $v^\flat$  by $v^\flat(u) = \la v,u\ra$ for all vectors $u$ at the point.  Given a covector $\alpha$ at a point,  define the vector $\alpha^\sharp$ by $\la \alpha^\sharp, u \ra = \alpha (u)$ for all tangents $u$.  In components, $(v^\flat)_i = g_{ij} v^j$ and $(\alpha^\sharp)^i = g^{ij} \alpha_j$, where $g^{ij}$ are the matrix elements for the inverse of the matrix with elements $g_{ij}$.  
For a vector field $B$, the components of $B^\flat$ are called the covariant components of $B$.  For a 1-form $\alpha$, the components of $\alpha^\sharp$ are called the contravariant components of $\alpha$.
Thus for example, $\int_\gamma B\cdot d\ell$ along a curve $\gamma$ is  $\int_\gamma B^\flat = \int_\gamma B_i dx^i$.

A notion that lies at the heart of this paper is the flux 2-form $\beta = i_B\Omega$ associated to any magnetic field $B$ by a volume-form $\Omega$. In a Riemannian manifold there is a natural volume-form up to sign (corresponding to a choice of orientation), namely $\Omega = \sqrt{|g|}\, dx^1 \wedge \cdots \wedge dx^d$, where $|g|$ denotes the determinant of the metric tensor $g_{ij}$. Thus, for $d=3$, the components of the flux form are given by $\beta_{ij}=\sqrt{|g|}\,\epsilon_{ijk}B^k$ (no summation), where $\epsilon_{ijk}$ is the Levi-Civita symbol. Note that they form a skew-symmetric $3\times3$ matrix, hence degenerate, which has rank 2 wherever $B$ does not vanish.

Another issue particularly relevant to this paper is the use of a vector potential for a divergence-free field $B$.  It is usual to consider the vector potential as a vector field $A$ such that $B= \curl\, A$, but far more natural to consider it as a 1-form $\alpha=A^\flat$ such that $\beta = d\alpha$. Then $$B^i = |g|^{-1/2}\, \epsilon^i_{jk} \frac{\partial \alpha_k}{\partial x^j},$$ where $\epsilon^i_{jk}$ is equal to $\pm 1$ if $ijk$ is an even or odd permutation of $123$, or $0$ if neither. Thus, the only place the metric enters this representation is via the prefactor $|g|^{-1/2}$, representing volume.  Contrast the formulae for $\curl$ in physical components!

\section{Magnetic fields as Hamiltonian systems}
\label{app:Hamsys}
It is commonplace that magnetic fieldline flow can be considered as a Hamiltonian system.  This is often done by assuming the field has a component, say $B^\phi$, of constant sign and then considering $B$ as a non-autonomous Hamiltonian system of one degree of freedom.  

A tidier way, in our opinion, is to think of it as an Arnol'd-Cartan Hamiltonian system.  These are defined on odd-dimensional manifolds by a closed 2-form $\beta$ with 1D kernel.  The trajectories are the curves whose tangent everywhere lies in $\ker \beta$.  The speed (and  direction) of motion along the curves is not defined, but a continuous choice can be made.  

The standard case is the dynamics of an autonomous Hamiltonian system $(M,\omega,H)$ on a regular energy level $H^{-1}(E)$, where $M$ is a manifold of even dimension $2n$, $\omega$ is a symplectic form, $H$ a smooth function, and $E\in \R$ is a regular value of $H$.  Then $\beta$ is the restriction $\omega_E$ of $\omega$ to $H^{-1}(E)$.  In this case there is a natural speed for the trajectories, namely that for $V$ defined by $i_V\omega = -dH$ on $M$.  It has the property that
\begin{equation}
i_V\mu_E = \omega_E^{\wedge (n-1)}/(n-1)!,
\label{eq:muE}
\end{equation}
where $\mu_E$ is the energy-surface-volume defined to be the restriction to $H^{-1}(E)$ of any $(2n-1)$-form $\mu$ on $M$ such that $\mu\wedge dH = \omega^{\wedge n}/n!$.  Indeed, given $\mu_E$, (\ref{eq:muE}) can be used
to choose the speed along $\ker \omega_E$.

Analogously, given a volume-form $\Omega$ in 3D and a closed 2-form $\beta$, a speed for $B$ is determined by $i_B\Omega = \beta$.

\end{appendices}

\bibliographystyle{unsrt}
\bibliography{ConvKAMbib}

\end{document}